\documentclass[aps,prd,floats,preprintnumbers,nofootinbib]{revtex4}
\usepackage{epsfig,graphicx,color,amsmath}

\begin{document}

\preprint{LA-UR-09-03593}
\preprint{NUHEP-TH/09-08}

\title{Pseudo-Dirac Neutrinos in the New Standard Model}

\author{Andr\'e de Gouv\^ea}
\affiliation{Northwestern University, Department of Physics \& Astronomy, 2145 Sheridan Road, Evanston, IL~60208, USA}

\author{Wei-Chih Huang}
\affiliation{Northwestern University, Department of Physics \& Astronomy, 2145 Sheridan Road, Evanston, IL~60208, USA}

\author{James Jenkins}
\affiliation{Elementary Particles and Field Theory Group, MS B285, Los Alamos National Laboratory, Los Alamos, NM 87545, USA}
\affiliation{Northwestern University, Department of Physics \& Astronomy, 2145 Sheridan Road, Evanston, IL~60208, USA}


\begin{abstract}
The addition of gauge singlet fermions to the Standard Model Lagrangian renders the neutrinos massive and allows one to explain all that is experimentally known about neutrino masses and lepton mixing by varying the values of the Majorana mass parameters $M$ for the gauge singlets and the neutrino Yukawa couplings $\lambda$. Here we explore the region of parameter space where $M$ values are much smaller than the neutrino Dirac masses $\lambda v$.  In this region, neutrinos are pseudo-Dirac fermions. We find that current solar data constrain $M$ values to be less than at least $10^{-9}$~eV, and discuss the sensitivity of future experiments to tiny gauge singlet fermion masses. We also discuss a useful basis for analyzing pseudo-Dirac neutrino mixing effects. In particular, we identify a simple relationship between elements of $M$ and the induced enlarged mixing matrix and new mass-squared differences. These allow one to directly relate bounds on the new mass-squared differences to bounds on the singlet fermion Majorana masses. 

\end{abstract}

\maketitle

\section{Introduction}
\label{sec:intro}

Nonzero neutrino masses reveal that the minimum standard model needs to be modified. Current data (mostly from solar, atmospheric, reactor and accelerator neutrino oscillation experiments), however, provide only minimal insight as to {\sl how} the standard model ought to be extended. Several completely different new physics scenarios can be constructed and all safely agree with observations, {\it i.e.}, all lead to small neutrino masses and non-trivial mixing among the three so-called active neutrino weak eigenstates, $\nu_e$, $\nu_{\mu}$, and $\nu_{\tau}$. Additions to the standard model that fit the data include  right-handed neutrinos, $SU(2)_L$ triplet fermions or scalars, lepto-quarks, etc. Recent summaries of the current experimental and theoretical situation can be found, for example, in \cite{GonzalezGarcia:2007ib,Valle:2006vb,Strumia:2006db,Mohapatra:2006gs,Mohapatra:2005wg,DeGouvea:2005gd,TASI}.

One version for the new standard model, $\nu$SM, consists of the minimum standard model augmented by a few (at least two) $SU(2)_L\times U(1)_Y$ gauge singlet Weyl fermions $N$, normally referred to as right-handed neutrinos. At the renormalizable level, the most general $\nu$SM Lagrangian consistent with  $SU(3)_c\times SU(2)_L\times U(1)_Y$ gauge invariance is
\begin{equation}
{\cal L}_{\nu\rm SM}={\cal L}_{\rm old}-\lambda_{\alpha i}L^{\alpha}HN^i-\sum_{i,j=1}^n\frac{M_{ij}}{2}N^iN^j + H.c.,
\label{l_new}
\end{equation}
where ${\cal L}_{\rm old}$ is the minimal standard model Lagrangian, $L$ are the $SU(2)_L$ lepton doublet fields $L=\left(\nu~\ell \right)^T$ and $H$ is the standard model Higgs doublet field. $\lambda_{\alpha i}$ are neutrino Yukawa couplings and $M_{ij}$ are Majorana masses for the $N$ fields. Note that $M$ is a symmetric matrix: $M_{ij}=M_{ji}$. $\alpha=e,\mu,\tau$,  $i=1,\ldots n$, and $n\ge 2$ is the number of right-handed neutrino fields.  After electroweak symmetry breaking, Eq.~(\ref{l_new}) describes $3+n$ Majorana neutral fermions, referred to as neutrinos. In general, all neutrino mass eigenstates $\nu_1,\nu_2,\ldots,\nu_{3+n}$ (with mass $m_1, m_2,\ldots,m_{3+n}$) are linear combinations of the active neutrinos plus the $n$ right-handed neutrinos. Since these singlet fermions do not couple to any of the standard model gauge bosons, they will also be referred to as sterile neutrinos. 

The phenomenology associated to Eq.~(\ref{l_new}) depends dramatically on the values of the $\nu$SM parameters $\lambda$ and $M$ (see, for example, \cite{deGouvea:2005er}). Even after the current data are taken into account, the spectrum of possibilities remains vast. When the matrix $M$ vanishes exactly, the six\footnote{For the remainder of this section, we restrict our discussion to $n=3$ right-handed neutrinos.} neutral states fuse into three Dirac fermions with masses proportional to the (square-roots of the) eigenvalues of $\lambda^{\dagger}\lambda$. In this case, the classical global symmetry of ${\cal L}_{\nu\rm SM}$ is enhanced and includes the lepton-number symmetry $U(1)_L$ (and its non-anomalous cousin $U(1)_{B-L}$). This indicates that $M$ can be interpreted as a symmetry breaking parameter (when $\lambda\neq 0$) so that (i) quantum corrections to $M$ are proportional to $M$ itself and hence (ii) {\sl any} value of $M$ is technically natural.

When $M\neq 0$, one can identify three qualitatively distinct regions of the $\lambda,M$ parameter space. In the region where $M\gg\lambda v$,\footnote{Beware of the abuse of notation and keep in mind that $\lambda$ and $M$ are matrices. For a more detailed discussion of this so-called seesaw limit, see, for example, \cite{deGouvea:2006gz}.} where $v$ is the vacuum expectation value of the neutral component of $H$, the six neutrino masses ``split'' into three lighter, mostly active states with masses generically of order $\lambda^2 v^2/M$ and three heavier, mostly sterile states with masses $M$. This phenomenon is referred to as the seesaw mechanism \cite{seesaw} and it was argued in \cite{deGouvea:2005er} that all values of $M\gtrsim 1$~eV are phenomenologically allowed. For small enough $M$ (in general $M$ much smaller than the weak scale), the $\nu$SM can be tested experimentally since the mostly sterile, heavier states are accessible at different facilities (see, for example, \cite{deGouvea:2006gz,Gorbunov:2007ak,Atre:2009rg}). It has also been pointed out that the mostly heavy states may qualify as viable warm dark matter \cite{Asaka:2005an}. Finally, for extraordinary choices of $\lambda$ (for recent discussions see \cite{de Gouvea:2007uz,Kersten:2007vk}), mostly sterile neutrinos with weak-scale masses can be detected at high energy collider experiments. 

The other two distinct regions of the $\lambda,M$ parameter space are characterized by $M\sim \lambda v$ and $M\ll \lambda v$. In the former, all six neutrino masses are of the same order, and all six neutrino mass eigenstates are characterized by ``homogeneous'' mixtures of active and sterile flavors. Such a scenario is both hard to study quantitatively and severely constrained by current solar and atmospheric neutrino data. In this paper, we explore the latter possibility: right-handed neutrino Majorana masses $M$ much {\sl smaller} than the so-called Dirac neutrino masses  $\lambda v$. Under these circumstances, neutrinos are pseudo-Dirac fermions \cite{pseudodirac,0nubb_pseudodirac,others}. 

Pseudo-Dirac neutrinos are Majorana neutrinos made up of roughly fifty-fifty mixtures of active and sterile neutrinos, and come in quasi-degenerate pairs. These will be properly defined in Sec.~\ref{sec:model}. In the limit $M\to 0$, neutrinos are Dirac fermions and hence can accommodate all current experimental data (as discussed above) so our main goal is to estimate an upper bound for $M$. Given our current understanding of neutrinos, the most stringent constraints on very small $M$ values are provided by solar data. The data, the estimation procedure, and our results are presented in Sec.~\ref{sec:solar}. Other constraints and predictions, as well as expectations for the future, are discussed in Sec.~\ref{sec:future}. A summary of what is currently known about pseudo-Dirac $\nu$SM neutrinos and the $\nu$SM Lagrangian (Eq.~(\ref{l_new})) in general is presented in Sec.~\ref{sec:conc}.

\setcounter{equation}{0} 
\setcounter{footnote}{0} 
\section{``Anti-Seesaw:'' Masses and Mixing}
\label{sec:model}

In the case of $n=3$ right-handed neutrinos, after electroweak symmetry breaking, the $6\times 6$ Majorana neutrino mass matrix is 
\begin{equation}
M_{\nu}=\left(\begin{array}{cc}0_3&m \\ m^T & \epsilon_m \end{array}\right),
\label{eq:Mnu}
\end{equation}
where $0_3$ stands for the $3\times 3$ zero matrix,  
\begin{equation}
\epsilon_m=\epsilon_m^T=V_R^T\epsilon_m^D V_R,
\end{equation}
is the symmetric $3\times 3$ Majorana right-handed neutrino mass matrix, and we choose the weak basis where the $3\times 3$ Dirac neutrino mass matrix is written as 
\begin{equation}
m=U^* m^D.
\end{equation}
This can be achieved by redefining the right-handed neutrino fields $N$. Above, the superscript $D$ indicates a diagonal matrix while $V_R$ and $U$ are unitary matrices. 

Since we are interested In the limit $m^D\gg \epsilon_m^D$ -- Dirac masses much larger than Majorana right-handed neutrino masses -- we can write
\begin{equation}
M_{\nu}\simeq \left(\begin{array}{cc}1_3&-\delta^* \\ \delta^T & 1_3\end{array}\right)
\frac{1}{\sqrt{2}}\left(\begin{array}{cc}U^*&-U^* \\ 1_3 & 1_3\end{array}\right)
\left(\begin{array}{cc}m^D(1_3+\epsilon^D)& 0_3 \\ 0_3 & -m^D(1_3-\epsilon^D)\end{array}\right)
\frac{1}{\sqrt{2}}\left(\begin{array}{cc}U^{\dagger}&1_3\\ -U^{\dagger} & 1_3\end{array}\right)
\left(\begin{array}{cc}1_3&\delta \\ -\delta^{\dagger} & 1_3\end{array}\right),
\label{Mnu_approx}
\end{equation}
where $1_3$ is the $3\times3$ unit matrix, and the elements of $\delta$ are small (this will be shown {\it a posteriori}) so that 
\begin{equation}
\left(\begin{array}{cc}1_3&\delta \\ -\delta^{\dagger} & 1_3\end{array}\right)\left(\begin{array}{cc}1_3&-\delta \\ \delta^{\dagger} & 1_3\end{array}\right) = \left(\begin{array}{cc}1_3&0_3 \\ 0_3 & 1_3\end{array}\right) + {\cal O}(\delta\delta^{\dagger},\delta^{\dagger}\delta).
\end{equation}
$\epsilon^D$ is a diagonal dimensionless matrix of small numbers.
Eq.~(\ref{Mnu_approx}) is satisfied, at leading order, if
\begin{equation}\delta=U\left(\frac{\epsilon^D}{2}+\varepsilon\right),
\end{equation}
where $\varepsilon$ is a dimensionless matrix of small numbers whose diagonal elements vanish. Both $\epsilon^D$ and $\varepsilon$ are functions of $\epsilon_m$ and $m^D$: 
\begin{eqnarray}
&\epsilon_m=2\epsilon^Dm^D+\varepsilon^T m^D+m^D\varepsilon, \label{em=ed,vare} \\
&{\rm and}~~m^D\varepsilon^T=-\varepsilon m^D .
\end{eqnarray}
 Note that $m^D$ and $\varepsilon$ do not commute and that $\epsilon_m$ is indeed a symmetric matrix with units of mass. Throughout we will ignore the possibility, ruled out by data, that different diagonal entries of $m^D$ are identical. According to Eq.~(\ref{em=ed,vare}), in the weak basis of choice and at leading order, the diagonal elements of $\epsilon_m$, proportional to $\epsilon^D$,  determine the mass-squared splittings between the quasi-degenerate states, while the off-diagonal elements proportional to $\varepsilon$ contribute only to the active-plus-sterile mixing matrix.

The $6\times 6$ neutrino mixing matrix is
\begin{eqnarray}
V=\left(\begin{array}{cccccc} V_{e1} & V_{e2} & V_{e3} & V_{e1'} & V_{e2'} & V_{e3'} \\
V_{\mu1} & V_{\mu2} & V_{\mu3} & V_{\mu1'} & V_{\mu2'} & V_{\mu3'} \\
V_{\tau1} & V_{\tau2} & V_{\tau3} & V_{\tau1'} & V_{\tau2'} & V_{\tau3'} \\
V_{s_11} & V_{s_12} & V_{s_13} & V_{s_11'} & V_{s_12'} & V_{s_13'} \\
V_{s_21} & V_{s_22} & V_{s_23} & V_{s_21'} & V_{s_22'} & V_{s_23'} \\
V_{s_31} & V_{s_32} & V_{s_33} & V_{s_31'} & V_{s_32'} & V_{s_33'} \\
 \end{array}\right)&=&\frac{1}{\sqrt{2}}\left(\begin{array}{cc}1_3&-\delta \\ \delta^{\dagger} & 1_3\end{array}\right)
\left(\begin{array}{cc}U&-U \\ 1_3 & 1_3\end{array}\right), \\
&=&\frac{1}{\sqrt{2}}\left(\begin{array}{cc}U\left(1_3-\frac{\epsilon^D}{2}-\varepsilon\right)&-U\left(1_3+\frac{\epsilon^D}{2}+\varepsilon\right) \\ 1_3+\frac{\epsilon^D}{2}-\varepsilon & 1_3-\frac{\epsilon^D}{2}+\varepsilon\end{array}\right), \label{V_2+1}
\end{eqnarray}
where we define the weak eigenstates as $\nu_{\alpha}$, $\alpha=e,\mu,\tau,s_1,s_2,s_3$ and the mass eigenstates as $\nu_i$, $i=1,2,3,1',2',3'$. In the Dirac limit $\epsilon_m\to 0$, $V_{\alpha i}=-V_{\alpha i'}=U_{\alpha i}/\sqrt{2}$ ($\alpha=e,\mu,\tau,~i=1,2,3$) and $m_i=-m_{i}'$ ($i=1,2,3$). We will order our states in the ``usual'' way \cite{Amsler:2008zzb}: $m_1^2<m_2^2<m_3^2$, $m_{1'}^2<m_{2'}^2<m_{3'}^2$ ($m_3^2<m_1^2<m_2^2$, $m_{3'}^2<m_{1'}^2<m_{2'}^2$) in the case of a normal (inverted) mass hierarchy. We assume the $\epsilon^D$ parameters small enough that the same mass hierarchy applies for the primed and unprimed eigenmasses. More specifically, $|m_{i}^2-m^2_{i'}|\ll m^2_{i}$ for all $i=1,2,3$.

In the next section we will concentrate on two subsets of the six neutrino mixing scenario: the case of two active and one right-handed neutrino, and the case of two active and two right-handed neutrinos. These will be described in more detail below and should serve as more concrete pedagogical examples.

In the case of two active ($\nu_e$ and $\nu_a$, a linear combination of $\nu_{\mu}$ and $\nu_{\tau}$) and one sterile neutrino $\nu_s$, the $3\times 3$ neutrino mass matrix can be written as
\begin{equation}
M_{\nu}=\left(
\begin{array}{ccc}
0 & 0 & m\sin\theta \\
0 & 0 & m\cos\theta \\
m\sin\theta & m\cos\theta & \epsilon_m \end{array}
\right), \label{M_2+1}
\end{equation}
where $\epsilon_m\equiv m\epsilon$ is the Majorana mass of the right-handed neutrino. In the limit $\epsilon\ll 1$ (and ignoring the case when $\theta$ is very close to 0 or $\pi/2$), and assuming all parameters are real,
\begin{equation}
V^{T}M_{\nu}V=\left(\begin{array}{ccc}
                0 & 0 & 0 \\
                0 & m\left(1+\frac{\epsilon}{2}\right) & 0 \\
                0 & 0 & -m\left(1-\frac{\epsilon}{2}\right)
              \end{array}
\right),
\end{equation}
where
\begin{equation}\label{V2+1}
V=
\left(\begin{array}{ccc}
                V_{e1} & V_{e2} & V_{e2'} \\
                V_{a1} & V_{a2} &  V_{a2'} \\
                V_{s1} &  V_{s2} & V_{s2'}
             \end{array}\right)
=\left(
\begin{array}{ccc}
\cos\theta & \frac{\sin\theta}{\sqrt{2}}\left(1-\frac{\epsilon}{4}\right)
& -\frac{\sin\theta}{\sqrt{2}}\left(1+\frac{\epsilon}{4}\right)  \\
- \sin\theta & \frac{\cos\theta}{\sqrt{2}}\left(1-\frac{\epsilon}{4}\right)
& -\frac{\cos\theta}{\sqrt{2}}\left(1+\frac{\epsilon}{4}\right) \\
0 &  \frac{1}{\sqrt{2}}\left(1+\frac{\epsilon}{4}\right) &
\frac{1}{\sqrt{2}}\left(1-\frac{\epsilon}{4}\right) \end{array}
\right).
\end{equation}
This system is described by three neutrino mass eigenstates: a massless one, $\nu_1$, which is a linear combination of $\nu_e$ and $\nu_a$, and two massive ones, $\nu_{2,2'}$, which are almost degenerate in mass-squared in the limit $\epsilon\ll 1$: $m_{2,2'}^2\simeq m^2(1\pm\epsilon)$, $m_2^2-m_{2'}^2=2m^2\epsilon$.  In vacuum, the survival probability of electron neutrinos $P_{ee}$ with energy $E$ after a distance $L$ has been traversed is
\begin{eqnarray}
1-P_{ee}&=&\frac{\sin^22\theta}{2}\left(1-\frac{\epsilon}{2}\right)\sin^2\left(\frac{m^2(1+\epsilon)L}{4E}\right)+\frac{\sin^22\theta}{2}\left(1+\frac{\epsilon}{2}\right)\sin^2\left(\frac{m^2(1-\epsilon)L}{4E}\right)+\sin^4\theta\sin^2\left(\frac{2m^2\epsilon L}{4E}\right), \nonumber \\
&=&\sin^22\theta\sin^2\left(\frac{m^2L}{4E}\right)+{\cal O}(\epsilon^2) \label{Pvac2+1}.
\end{eqnarray}
On the other hand, the oscillation probabilities of active neutrinos into sterile neutrinos are
\begin{eqnarray}
P_{es}=\sin^2\theta\sin^2\left(\frac{2m^2\epsilon L}{4E}\right), \\
P_{as}=\cos^2\theta\sin^2\left(\frac{2m^2\epsilon L}{4E}\right). 
\end{eqnarray}

Eq.~(\ref{Pvac2+1}) agrees, of course,  with the well-known two-neutrino oscillation probability in vacuum in the limit $\epsilon\ll 1$. On the other hand, in the very long distance limit, $2m^2\epsilon L\sim E$, the oscillation due to the small mass-squared splitting can be observed on top of the averaged-out ``active--active'' oscillations, 
\begin{equation}
\lim_{L\gg \frac{E}{m^2}}\left(1-P_{ee}\right)=\frac{\sin^22\theta}{2}+\sin^4\theta\sin^2\left(\frac{2m^2\epsilon L}{4E}\right).
\end{equation}

In the case of two active ($\nu_e$ and $\nu_a$, a linear combination of $\nu_{\mu}$ and $\nu_{\tau}$) and two sterile neutrinos $\nu_{s_1}$ and $\nu_{s_2}$, the $4\times 4$ neutrino mass matrix can be expressed as follows, assuming all parameters real:
\begin{equation}\label{M_2+2}
M_{\nu}=\left(
   \begin{array}{cccc}
     0 & 0 & m_1\cos\theta  & m_2\sin\theta \\
     0 & 0 & -m_1\sin\theta & m_2\cos\theta \\
     m_1\cos\theta & -m_1\sin\theta & \epsilon_{m1} & \epsilon_{m3} \\
     m_2\sin\theta & m_2\cos\theta & \epsilon_{m3} & \epsilon_{m2} \\
   \end{array}
 \right).
\end{equation}
For concreteness, we allow $\theta\in[0,\pi/2]$ and define $m_2^2>m_1^2$. In this case, re-expressing $\epsilon_{m1}=m_1\epsilon_1$, $\epsilon_{m2}=m_2\epsilon_2$, $\epsilon_{m3}=[(m_2^2-m_1^2)/m_2]\epsilon_3$, and assuming all $\epsilon_i\ll 1$ and $\theta$ not too close to 0 or $\pi/2$,
\begin{equation}
V^{T}M_{\nu}V=\left(\begin{array}{cccc}
                m_1\left(1+\frac{\epsilon_1}{2}\right) & 0 & 0 & 0\\
                0 & m_2\left(1+\frac{\epsilon_2}{2}\right) & 0 & 0 \\
                0 & 0 & -m_1\left(1-\frac{\epsilon_1}{2}\right) & 0 \\
                0 & 0 & 0 & -m_2\left(1-\frac{\epsilon_2}{2} \right)
              \end{array}
\right),
\end{equation}
where
\begin{equation}\label{V4x4}
V=
\left(\begin{array}{cccc}
                 V_{e1} & V_{e1'} & V_{e2} & V_{e2'} \\
                 V_{a1} & V_{a1'} & V_{a2} &  V_{a2'} \\
                 V_{s_11} &  V_{s_11'} & V_{s_12} & V_{s_12'} \\
                 V_{s_21} &  V_{s_21'} & V_{s_22} & V_{s_22'} \\
              \end{array}\right)
=\frac{1}{\sqrt{2}}\left(
\begin{array}{cc}
U_2(1-E)_2 & -U_2(1+E)_2   \\
(1+E)_2^{\dagger} & (1-E)_2^{\dagger}  \end{array}
\right),
\end{equation}
and the $2\times 2$ matrices
\begin{equation}
U_2=\left(
\begin{array}{cc}
\cos\theta & \sin\theta   \\
-\sin\theta & \cos\theta \end{array}
\right), ~~~
(1\pm E)_2=\left(
\begin{array}{cc}
1\pm\frac{\epsilon_1}{4} & \mp\frac{m_1}{m_2}\epsilon_3   \\
\pm\epsilon_3 & 1\pm\frac{\epsilon_2}{4} \end{array}
\right). \label{U_2}
\end{equation}

This system is characterized by four mass eigenstates $\nu_1,\nu_2,\nu_{1'},\nu_{2'}$ which are pair-wise quasi-degenerate in mass-squared. Oscillations are described by six distinct oscillation frequencies: $\Delta m^2_{12}=m_2^2-m_1^2+m_2^2\epsilon_2-m_1^2\epsilon_1$, $\Delta m^2_{1'2}=m_2^2-m_1^2+m_2^2\epsilon_2+m_1^2\epsilon_1$, $\Delta m^2_{12'}=m_2^2-m_1^2-m_2^2\epsilon_2-m_1^2\epsilon_1$, $\Delta m^2_{1'2'}=m_2^2-m_1^2-m^2_2\epsilon_2+m_1^2\epsilon_1$, $\Delta m^2_{1'1}=2m_1^2\epsilon_1$, and  $\Delta m^2_{2'2}=2m_2^2\epsilon_2$ (in the limit $\epsilon_1,\epsilon_2\ll 1$). It is convenient to also express the two small frequencies in terms of the elements of the Majorana mass matrix for the right-handed neutrinos: $\Delta m^2_{1'1}=2m_1\epsilon_{m1}$, and  $\Delta m^2_{2'2}=2m_2\epsilon_{m2}$. If all elements of $\epsilon_m$ are of the same order of magnitude, one expects $\Delta m^2_{2'2}>\Delta m^2_{1'1}$ since $m_2^2>m_1^2$.

In vacuum, the survival probability of electron neutrinos $P_{ee}$ with energy $E$ after a distance $L$ has been traversed is 
\begin{eqnarray}
1-P_{ee}&=&\frac{\sin^22\theta}{4}\left[\left(1-\frac{\epsilon_1}{2}-\frac{\epsilon_2}{2}\right)\sin^2\left(\frac{\Delta
m^2_{12}L}{4E}\right)+\left(1-\frac{\epsilon_1}{2}+\frac{\epsilon_2}{2}\right)\sin^2\left(\frac{\Delta
m^2_{1'2}L}{4E}\right)+\right. \nonumber \\
&& \left.\left(1+\frac{\epsilon_1}{2}-\frac{\epsilon_2}{2}\right)\sin^2\left(\frac{\Delta
m^2_{12'}L}{4E}\right)+\left(1+\frac{\epsilon_1}{2}+\frac{\epsilon_2}{2}\right)\sin^2\left(\frac{\Delta
m^2_{1'2'}L}{4E}\right)\right] +\nonumber \\
&& + \epsilon_3\frac{\sin2\theta}{2m_2}\left\{(\cos^2\theta
m_1-\sin^2\theta m_2)\cos\left(\frac{\Delta m^2_{12} L}{2E}\right)+
(\cos^2\theta m_1+\sin^2\theta m_2)\cos\left(\frac{(\Delta
m^2_{12}-\Delta m^2_{11'}) L}{2E}\right) \right. \nonumber\\
&&+\left. (-\cos^2\theta m_1+\sin^2\theta m_2)\cos\left(\frac{(\Delta
m^2_{11'}-\Delta m^2_{12'}) L}{2E}\right)-
(\cos^2\theta m_1+\sin^2\theta m_2)\cos\left(\frac{\Delta m^2_{12'}
L}{2E}\right)
\right\} \nonumber \\
&&+\cos^4\theta\sin^2\left(\frac{\Delta
m^2_{1'1}L}{4E}\right)+\sin^4\theta\sin^2\left(\frac{\Delta
m^2_{2'2}L}{4E}\right), \\
&=& \sin^22\theta\sin^2\left(\frac{(m_2^2-m_1^2)L}{4E}\right)+{\cal
O}(\epsilon^2)
\label{Pvac2+2}.
\end{eqnarray}
Above, ${\cal O}(\epsilon^2)$ indicates terms which are of order the product of two $\epsilon_i$, $i=1,2,3$ ($\epsilon_1^2,\epsilon_1\epsilon_2$, etc). It is trivial to note that, for small enough $\epsilon_1,\epsilon_2,\epsilon_3$ the two-flavor vacuum oscillation expression is reproduced, as expected. Finally, in the very long baseline limit, assuming that the fast $12$ (and $1'2, 12', 1'2'$) oscillations average out, 
\begin{equation}
\lim_{L\gg \frac{E}{m^2_2-m_1^2}}(1-P_{ee})=\frac{\sin^22\theta}{2}+\cos^4\theta\sin^2\left(\frac{2m_1^2\epsilon_1L}{4E}\right)+\sin^4\theta\sin^2\left(\frac{2m_2^2\epsilon_2L}{4E}\right).
\label{long_2x2}
\end{equation} 
Again as expected, in the very long-baseline limit the electron neutrino survival probability is equal to the averaged out ``active oscillation'' effect, plus two long-wavelength components driven by the $1'1$ and $2'2$ mass-squared splittings. Two features are readily visible: one is that there no dependency on the off-diagonal $\epsilon_3$ parameter (this only appears as ${\cal O}(\epsilon^2)$ level corrections to the coefficients of the different terms in Eq.~(\ref{long_2x2})). The second is that, assuming $\cos^4\theta\sim\sin^4\theta$, we expect to be more sensitive to $\epsilon_2$ than $\epsilon_1$, as the $2'2$-oscillations ``turn on'' before the $1'1$-oscillations (remember $m_2^2>m_1^2$).
 
\setcounter{equation}{0}  
\setcounter{footnote}{0} 
\section{Current Constraints}
\label{sec:solar}

If one assumes that there are three right-handed neutrinos with Majorana masses much smaller than the Dirac neutrino masses that govern ``active'' neutrino oscillations, current experimental data can be used to constrain the right-handed neutrino Majorana mass matrix  $\epsilon_m$. There are, however, too many free parameters in $\epsilon_m$ and the individual upper bound on each of these is not very illuminating and outside the aspirations of this work. Such a bound will also depend on aspects of neutrino masses and mixing that are currently unknown, including the neutrino mass hierarchy and $|U_{e3}|^2$. We are, however, interested in asking what is the constraint on $\epsilon_m$ assuming that all its elements are of the same order of magnitude. Hence we will concentrate on the current experimental upper bound on the element of $\epsilon_m$ that is best constrained. This will become clear in the following paragraphs.

If the three neutrino mass eigenstates $\nu_1,\nu_2,\nu_3$ identified experimentally  \cite{Amsler:2008zzb} are in reality ``split'' into, say,\footnote{The current data are also consistent with two right-handed neutrinos. In this case, the massless neutrino mass eigenstate is not split, and also does not contain a sterile neutrino component. This is identical to the 2+1 case [Eq.~(\ref{M_2+1})] discussed in the previous section.} six states $\nu_1,\nu_2,\nu_3,\nu_{1'},\nu_{2'},\nu_{3'}$. As discussed in the previous section, these splittings will manifest themselves via new, very long wavelength oscillations characterized by the $\Delta m^2_{i'i}$ mass-squared differences ($i=1,2,3$). These, on the other hand, are proportional to $\epsilon_{m_{ii}}m_i$ and the largest new mass-squared splitting is associated to the $\epsilon_{m_{ii}}$ value associated to the largest $m_i^2$. In turn, the largest $m_{i}^2$ value depends on the neutrino mass hierarchy. In the case of an inverted mass hierarchy, $m_2^2>m_1^2 \gtrsim 2\times 10^{-3}$~eV$^2$ is the largest $m_i^2$. On the other hand, if the hierarchy is normal, $m_3^2\gtrsim 2\times 10^{-3}$~eV$^2$ is the largest $m_i^2$ while $m_2^2>m_1^2$ and $m_2^2 \gtrsim 8\times 10^{-5}$~eV$^2$. 

Ignoring $|U_{e3}|$-driven effects, $\Delta m^2_{3'3}$ is best constrained by the disappearance of muon-type neutrinos and antineutrinos produced in the atmosphere. We estimate that these experiments are sensitive to $\epsilon_m$ values that lead to new oscillation lengths which are not much larger than the diameter of the earth, or
\begin{equation}
L_{\rm osc}^{3'3}=\frac{1}{\epsilon_3}\left(\frac{\rm 10^{-3}~eV^2}{m_3^2}\right)\left(\frac{E}{\rm 100~MeV}\right)\times 10^2~\rm km \lesssim 10^4~km.
\end{equation}
Hence atmospheric experiments can ``see'' $\epsilon_3$ values larger than around $10^{-2}$. This, in turn, translates into a sensitivity to $\epsilon_m$ elements of order $\epsilon_{m}\gtrsim\sqrt{10^{-3}}\times10^{-2}$~eV. A more detailed estimate can be extract from the analyses performed in \cite{Cirelli:2004cz}. In more detail, atmospheric data constrain the new mass-squared difference to be less than about $10^{-4}$~eV$^2$. In the case of a normal mass hierarchy, this translates into $4\epsilon_3\times 10^{-3}~{\rm eV^2}\lesssim 10^{-4}~{\rm eV^2}$ or $\epsilon_3\lesssim 0.03$ and $\epsilon_m\lesssim 0.001$~eV. The estimate above was made for $m_3^2=2\times 10^{-3}$~eV$^2$.

$\Delta m^2_{2'2}$ (and $\Delta m^2_{1'1}$) is best constrained by solar neutrino experiments. Ultimately, one is sensitive to oscillation lengths of order the earth--sun distance (1~A.U.$=149.6\times 10^{6}$~km). Naively, the sensitivity to $\epsilon_2$ and $\epsilon_1$ can be estimated from
\begin{equation}
L_{\rm osc}^{1'1,2'2}=\frac{10^{-7}}{\epsilon_{1,2}}\left(\frac{\rm 8\times 10^{-5}~eV^2}{m_{1,2}^2}\right)\left(\frac{E}{\rm 1~MeV}\right)\times~1~\rm A.U. \lesssim 1~A.U.,
\end{equation}
which translates into $\epsilon_{1,2}\gtrsim 10^{-7}$ if both $m_1$ and $m_2$ are of the same order. This translates into a sensitivity to $\epsilon_m$ elements of order $\epsilon_m\gtrsim\sqrt{8\times 10^{-5}}\times 10^{-7}$~eV. Both sensitivity estimates ($\epsilon_3\gtrsim 10^{-2}$ and $\epsilon_{1,2}\gtrsim 10^{-7}$) are quoted assuming the Dirac neutrino mass hierarchy is normal. In the case of an inverted hierarchy the sensitivity of atmospheric neutrino experiments to $\epsilon_3$ is markedly worse, while that of solar neutrino experiments to $\epsilon_{1,2}$ is markedly better. Finally, if the Dirac neutrino masses are quasi-degenerate and much larger in magnitude than $\sqrt{\Delta m^2_{13}}\sim 0.05$~eV (the atmospheric mass-squared difference), the sensitivity to all $\epsilon_i$ is expected to be better than estimated above. 

In conclusion, constraints on $\epsilon_{1,2}$ from solar neutrino data provide, by far, the best bound on a given $\epsilon_m$ element, regardless of the neutrino mass hierarchy. Therefore, we will concentrate on the effect of the seesaw right-handed neutrinos in solar oscillations. We will further assume $|U_{e3}|$ is small enough so that $\nu_3$ and $\nu_{3'}$ related effects in experiments with electron neutrinos in the initial state are negligible and will concentrate on an effective system consisting of the electron neutrino, the linear combination $\nu_a$ of $\nu_{\mu}$ and $\nu_{\tau}$ orthogonal to the one in $\nu_{3,3'}$ and either one or two sterile states. 

\subsection{Oscillation of Solar Neutrinos,  2+1 case}

We first discuss the case of only one right-handed neutrino, Eq.~(\ref{M_2+1}), where $m_1\equiv 0$. The $2+1$ scenario captures most of the physics of the 2+2 case, which will be discussed in the next subsection, and contains all the relevant information in the limit $m_1\ll m_2$. In the $2+1$ case, the effect of right-handed neutrinos is entirely captured by one dimensionless parameter $\epsilon$. Furthermore, the neutrino mass responsible for the oscillation of solar neutrinos and reactor antineutrinos at KamLAND is uniquely determined: $m^2=(7.59\pm0.21)\times 10^{-5}$~eV$^2$ \cite{:2008ee} from KamLAND data. We will show that $\epsilon$ is constrained to be small enough that this measurement is not affected by the presence of the sterile neutrino (see Eq.~(\ref{Pvac2+1})). The other parameter in $M_{\nu}$, the mixing angle $\theta$, is mostly determined by solar data, more specifically those from SNO and Super-Kamiokande. We will argue later that the impact of $\epsilon_m$ on the determination of $\theta$ is not significant. 

We first briefly discuss how neutrinos produced in  the core of
the sun propagate towards the surface of the sun and then to a detector on earth, and calculate the relevant transition probabilities.  
We  hence construct the effective Hamiltonian $H$ in the flavor basis in the presence of the matter  \cite{MSW}:
\begin{equation}
H=\frac{M^{\dagger}_{\nu}M_{\nu}}{2E}+\left(\begin{array}{ccc} A & & \\ & 0 & \\ & & 0 \end{array}\right)=\left(
      \begin{array}{ccc}
        A+\frac{m^2\sin^2\theta}{2E} & \frac{m^2\sin\theta\cos\theta}{2E} & \frac{m^2\epsilon\sin\theta}{2E} \\
        \frac{m^2\sin\theta\cos\theta}{2E} & \frac{m^2\cos^2\theta}{2E} & \frac{m^2\epsilon\cos\theta}{2E} \\
        \frac{m^2\epsilon\sin\theta}{2E} & \frac{m^2\epsilon\cos\theta}{2E} & \frac{m^2(1+\epsilon^2)}{2E} \\
      \end{array}
    \right), \label{H2+1}
    \end{equation}
where $A=\sqrt{2}G_{F}N_{e}$ is the matter potential due to electrons, $G_{F}$ is the Fermi constant, $N_e$ is the position-dependent electron number density, and $E$ is the neutrino energy. In principle, we should also consider the effect of neutral current interactions due to the presence of neutrons. This contribution, however, is expected to be negligible in the sun given that it consists mostly of hydrogen (hence ``neutron-poor''), and the proton and electron neutral current contributions cancel out. Under this approximation, analytical solutions for the transition probability can be easily obtained as long as $N_e$ is a simple enough function of the neutrino position. 

It is easy to understand how an electron neutrino produced in the sun's core propagates first to the sun's surface and later to the detectors on earth. It is illustrative to discuss what happens in the $\epsilon\to0$ limit. In this case, one of the eigenvectors of $H$ (Eq.~(\ref{H2+1})), with eigenvalue $m^2/2E$, is independent of $A$ and is purely sterile. In the basis defined by Eq.~(\ref{V2+1}) it is neither $\nu_2$ nor $\nu_{2'}$ but $1/\sqrt{2}(\nu_2+\nu_{2'})$.\footnote{In the limit $\epsilon\to 0$, $\nu_2$ and $\nu_{2'}$ have the same mass-squared and hence any linear combination of them is an eigenstate of the propagation Hamiltonian in vacuum.} The other two states are purely active and can be obtained by diagonalizing the familiar matter-affected two-by-two neutrino propagation Hamiltonian. For $m^2$ values of interest, electron neutrinos are produced in the sun's core and propagate adiabatically, exiting the sun as an incoherent mixture of $\nu_1$ and $1/\sqrt{2}(\nu_2-\nu_{2'})$ with probabilities $\cos^2\theta_M$ and $\sin^2\theta_M$, respectively, where the matter mixing angle is defined at the production region by the familiar expression
\begin{equation}
\sin2\theta_M=\frac{\Delta\sin2\theta}{\sqrt{(\Delta\cos2\theta-A)^2+\Delta^2\sin^22\theta}}, ~~~\cos2\theta_M=\frac{\Delta\cos2\theta-A}{\sqrt{(\Delta\cos2\theta-A)^2+\Delta^2\sin^22\theta}}, 
\label{theta_M}
\end{equation}
where $\Delta\equiv m^2/2E$. In this case, $P_{ee}$ at the surface of the sun (or the earth) is given by 
\begin{equation}
\lim_{\epsilon\to0}P_{ee}^{\rm solar}= \cos^2\theta_M|V_{e1}|^2 + \frac{\sin^2\theta_M}{2}\left|V_{e2}-V_{e2'}\right|^2=\cos^2\theta_M\cos^2\theta+\sin^2\theta_M\sin^2\theta,
\end{equation}
which is the familiar result for solar neutrino oscillations in the so-called LMA region.

\begin{figure}
\includegraphics[width=0.9\textwidth]{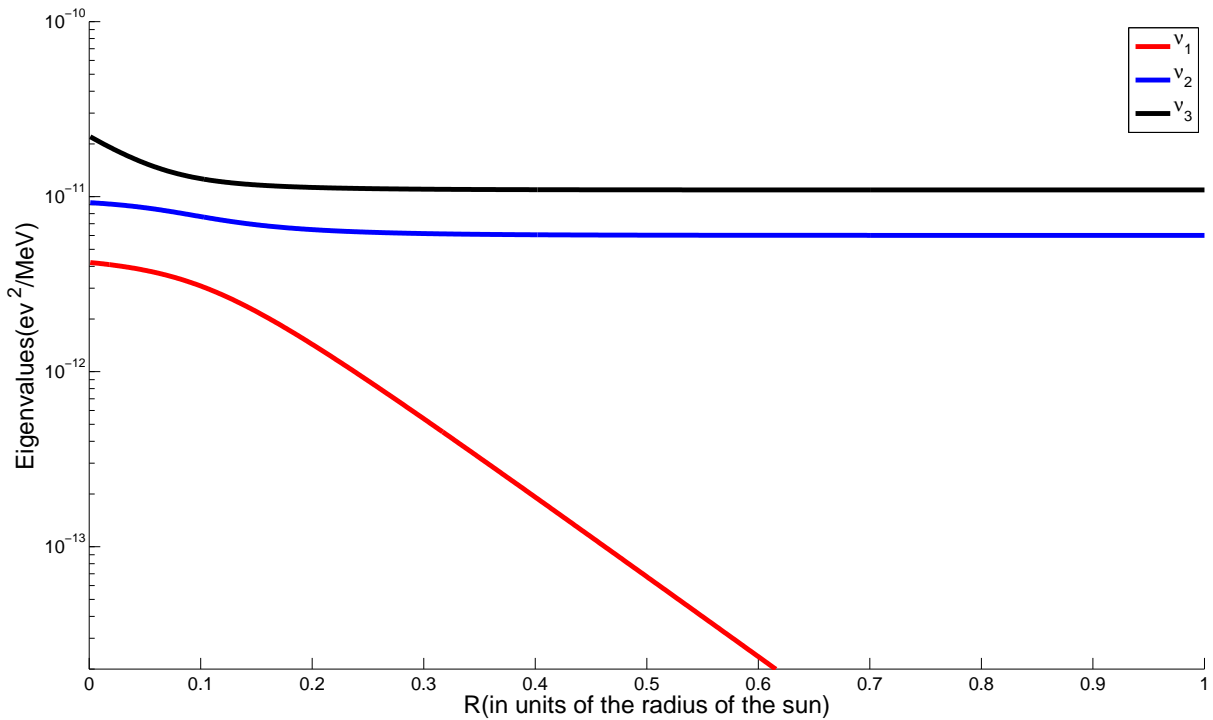}
\caption{The evolution of instantaneous eigenvalues of the Hamiltonian inside the sun  in the 2+1 scenario, Eq.~(\ref{H2+1}), for $E=5$~MeV,
$m^2=8.1\times 10^{-5}$~eV$^2$,
$\sin^2\theta=0.3$,
$\epsilon=3\times 10^{-1}$.}
\label{figa}
\end{figure}

Figure \ref{figa} depicts the evolution of the instantaneous eigenvalues of $H$ inside the sun, as a function $R$, the distance from the sun's center, for $\epsilon\neq 0$. For illustrative purposes we choose, for the Hamiltonian parameters, $E=5$~MeV,
$m^2=8.1\times 10^{-5}$~eV$^2$,
$\sin^2\theta=0.3$,
$\epsilon=3\times 10^{-1}$. The LMA MSW resonance (between the largest eigenvalue and the smallest one) can be readily identified (at $R\sim0.1$). One can also see the impact of $\epsilon\neq 0$: the top two eigenvalues are ``split'' for all values of  $R$. Furthermore, there is another ``resonance'' \cite{Alex} due to the small but non-zero $\epsilon$ at larger $R$ between the two heaviest Hamiltonian eigenstates. This will be discussed momentarily.

Assuming that the crossing of the LMA MSW resonance is adiabatic and that $\epsilon\ll 1$, the electron neutrino survival probability can be written as
\begin{eqnarray}
  P_{ee} &=&\left|
    \left(
      \begin{array}{ccc}
        1 & 0 & 0 \\
      \end{array}
    \right)
    V
    \left(
    \begin{array}{ccc}
            \exp(-i\phi^{'}_1) & 0 & 0 \\
            0 & \exp(-i\phi^{'}_2) & 0 \\
            0 & 0 & \exp(-i\phi^{'}_3) \\
   \end{array}
   \right)
    \left(
      \begin{array}{ccc}
       1 & 0 & 0 \\
        0 & \sqrt{1-P_c} & -\sqrt{P_c} \\
        0 & \sqrt{P_c} & \sqrt{1-P_c} \\
      \end{array}
    \right) \times \right. \notag \\
   & & \left.\left(
          \begin{array}{ccc}
            \exp(-i\phi_1) & 0 & 0 \\
            0 & \exp(-i\phi_2) & 0 \\
            0 & 0 & \exp(-i\phi_3) \\
          \end{array}
        \right)
        V_{\rm mat}^{\dagger}
    \left(
      \begin{array}{c}
        1 \\
        0 \\
        0 \\
      \end{array}
    \right)\right|^2, \label{P2+1}
\end{eqnarray}
where  \cite{Petcov_pc}
\begin{equation}
    P_c=\frac{e^{-\gamma|V_{s2}|^2}-e^{-\gamma}}{1-e^{-\gamma}},~~~~
    \gamma\simeq9.8\left(\frac{\epsilon}{10^{-4}}\right)\left(\frac{m^2}{8\times10^{-5}~\rm eV^2}\right)
               \left(\frac{\rm 0.862~MeV}{E}\right), \label{Pc}
\end{equation}
$V$ is given by Eq.~(\ref{V2+1}) and $V_{\rm mat}$ is the unitary matrix that diagonalizes Eq.~(\ref{H2+1}) at the production point. Eq.~(\ref{P2+1}) can be understood as follows. The electron neutrino is first expressed in the basis of the Hamiltonian at the production point. It then propagates adiabatically (each component acquiring a phase factor $\phi_i$, $i=1,2,3$) until it gets to the location of the resonance between the two ``heaviest'' instantaneous Hamiltonian eigenstates. Around that point, the unitary evolution is characterized by the matrix containing $P_c$, which is the crossing probability between the two Hamiltonian eigenstates. Finally, it propagates from that point until it gets to the earth (each component acquiring a phase factor $\phi'_i$, $i=1,2,3$), where we compute the probability that this state is an electron neutrino. For a similar discussion of $P_{ee}$ in the presence of a new small mass-squared difference, see, for example, \cite{Cirelli:2004cz}.

We can estimate $P_{ee}$ for $\epsilon\gtrsim10^{-3}$ when $P_c$ in Eq.~(\ref{Pc}) vanishes to a very good approximation for all neutrino energies below 10~MeV. If we consider the case $A\gg \Delta$ in the production region (this is an excellent approximation in the upper energy range of the $^8$B solar neutrino spectrum), $P_{ee}$ is very easy to compute since the electron neutrino is a Hamiltonian eigenstate at birth (corresponding to the ``heaviest'' Hamiltonian eigenstate) and the entire evolution of the state inside the sun is adiabatic. In summary, the electron neutrino exits the sun as the heaviest neutrino mass eigenstate, which we will assume to be $\nu_2$ (which is the case for $\epsilon$ positive) and
\begin{eqnarray}
&P_{ee}=|V_{e2}|^2=\frac{\sin^2\theta}{2}\left(1-\frac{\epsilon}{2}\right), \\
&P_{ea}=|V_{\mu2}|^2=\frac{\cos^2\theta}{2}\left(1-\frac{\epsilon}{2}\right), \\
&P_{es}=|V_{s2}|^2=\frac{1}{2}\left(1+\frac{\epsilon}{2}\right). \label{Pes_large_e}
\end{eqnarray}
It is easy to see that under these circumstances there is no value of $\theta$ that provides a good fit to the solar neutrino data, as we will discuss in more detail in Sec.~\ref{sub:exp}. 

\begin{figure}
\includegraphics[width=0.6\textwidth]{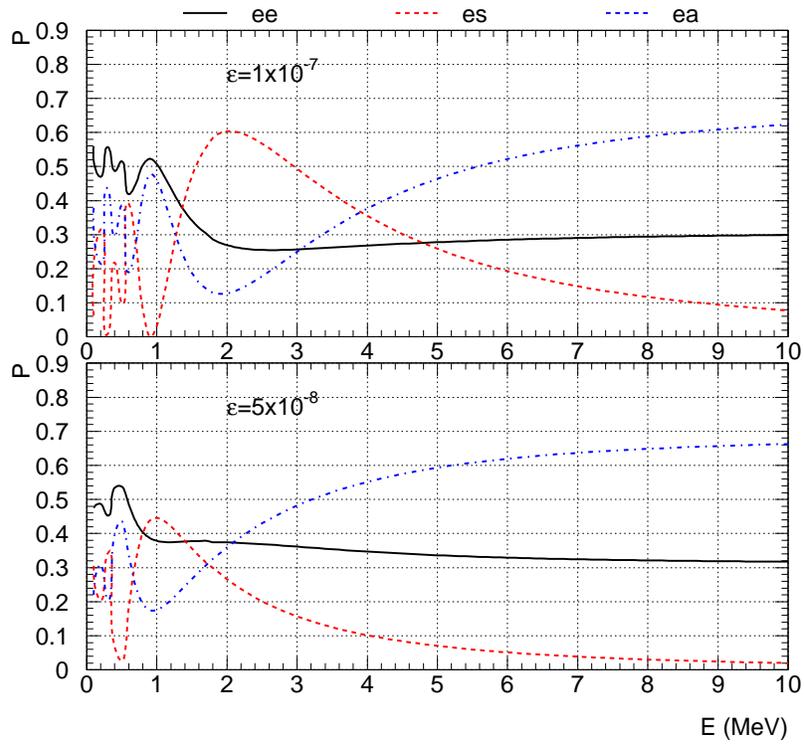}
\caption{$P_{e\zeta}$ for solar neutrinos at the surface of the earth, $\zeta=e,a,s$, as a function of neutrino energy for $m^2=7.6\times10^{-5}$~eV$^2$, $\sin^2\theta=0.31$. TOP:  $\epsilon=1\times 10^{-7}$, BOTTOM:  $\epsilon=5\times 10^{-8}$.}
\label{fig:P_2+1}
\end{figure}

On the other hand,  in the limit $\epsilon\ll 10^{-3}$,  it is sufficient to keep only the leading order ($\epsilon\to 0$) limit of $V_{\rm mat}$. In more detail, 
\begin{equation}
V_{\rm mat}^{\dagger}
    \left(
      \begin{array}{c}
        1 \\
        0 \\
        0 \\
      \end{array}
    \right)= \left(
      \begin{array}{c}
        \cos\theta_M \\
         \sin\theta_M \\
         0 \\
      \end{array}
    \right) + {\cal O}(\epsilon) \label{V_matter}
\end{equation}
where $\theta_M$ is defined in Eq.~(\ref{theta_M}). At the surface of the sun, after integrating over the neutrino production region, the electron neutrino is an incoherent mixture of $\nu_1$ (with probability $\cos^2\theta_M$) and $\sqrt{1-P_c}\nu_2-\sqrt{P_c}\nu_{2'}$\footnote{An irrelevant relative phase factor between $\nu_2$ and $\nu_{2'}$ has been omitted.} (with probability $\sin^2\theta_M$). The latter may undergo vacuum oscillations on its way between the sun's surface and the earth with an oscillation probability given by
\begin{eqnarray}
P(\sqrt{1-P_c}\nu_2-\sqrt{P_c}\nu_{2'}\to\nu_e)&=&\left|V_{e2}\sqrt{1-P_c}-V_{e2'}\sqrt{P_c}\exp\left[-i\frac{2m^2\epsilon L}{2E}\right]\right|^2, \nonumber \\
&=&\frac{\sin^2\theta}{2}\left[1+2\sqrt{P_c(1-P_c)}\cos\left(\frac{2m^2\epsilon L}{2E}\right)\right]. \label{Pmix_e}
\end{eqnarray}
Combining all the information above, 
\begin{eqnarray}
P_{ee}&=&\cos^2\theta_M\cos^2\theta+\sin^2\theta_M\frac{\sin^2\theta}{2}\left[1+2\sqrt{P_c(1-P_c)}\cos\left(\frac{2m^2\epsilon L}{2E}\right)\right], \nonumber \\
P_{ea}&=&\cos^2\theta_M\sin^2\theta+\sin^2\theta_M\frac{\cos^2\theta}{2}\left[1+2\sqrt{P_c(1-P_c)}\cos\left(\frac{2m^2\epsilon L}{2E}\right)\right], \label{P_sun_2+1} \\
P_{es}&=&\sin^2\theta_M\frac{1}{2}\left[1-2\sqrt{P_c(1-P_c)}\cos\left(\frac{2m^2\epsilon L}{2E}\right)\right]. \nonumber
\end{eqnarray}
Figure~\ref{fig:P_2+1} depicts $P_{e\zeta}$ $\zeta=e,a,s$ for solar neutrinos as a function of energy for $m^2=7.6\times10^{-5}$~eV$^2$, $\sin^2\theta=0.31$ and two different values of $\epsilon=1\times 10^{-7}$ and $\epsilon=5\times 10^{-8}$. In order to compute $P_{e\zeta}$, we integrate over the neutrino production region in the sun's core, taking into account that $\theta_M$ depends on $N_e$ at the production point. For high energy solar neutrinos $P_{es}$ is small and $P_{ee}$ is very similar to the standard LMA solution to the solar neutrino puzzle. For low energy solar neutrinos, the oscillatory pattern becomes more pronounced and $P_{es}$ becomes significantly nonzero.  As already briefly discussed, for much larger values of $\epsilon$, $P_c\to 0$ and $P_{es}\simeq\sin^2\theta_M/2$ is roughly equal to one half for high energy solar neutrinos. On the flip side, for smaller values of $\epsilon$, $P_c\to 1/2$ and the argument of the cosine in Eqs.~(\ref{P_sun_2+1}) is vanishingly small. In this case, as already discussed, sterile neutrino effects disappear and we recover the standard LMA solution to the solar neutrino puzzle.

\subsection{Oscillation of Solar Neutrinos,  2+2 case}

It is straight forward to generalize the results obtained in the previous subsection to the case two active ($\nu_e$ and $\nu_a$) and two sterile neutrinos ($\nu_1$ and $\nu_2$). In this case, the Hamiltonian that governs neutrino oscillations in neutral hydrogen matter characterized by an electron number density $N_e$ is 
\begin{equation}
  H = \frac{M_{\nu}^{\dagger}M_{\nu}}{2E}+\left(\begin{array}{cccc} A &&& \\ & 0 && \\ && 0 & \\ &&& 0 \end{array}\right) = \left(\begin{array}{cc}  U_2 & \\  & 1_2  \end{array}\right) 
  \left(\begin{array}{cccc}
  A\cos^2\theta+\frac{m_1^2}{2E} & A\cos\theta\sin\theta & \frac{\epsilon_1 m_1^2}{2E} & \frac{\epsilon_3 \left(\frac{\Delta m^2 m_1}{m_2}\right)}{2E} \\
  A\cos\theta\sin\theta &  A\sin^2\theta+\frac{m_2^2}{2E} &  \frac{\epsilon_3 \Delta m^2}{2E} & \frac{\epsilon_2 m_2^2}{2E} \\
                          \frac{\epsilon_1 m_1^2}{2E} & \frac{\epsilon_3 \Delta m^2}{2E} & \frac{m_1^2(1+{\cal O}(\epsilon^2)) }{2E} & {\cal O}(\epsilon^2) \\
                          \frac{\epsilon_3  \left(\frac{\Delta m^2m_1}{m_2}\right)}{2E} & \frac{\epsilon_2 m_2^2}{2E} & {\cal O}(\epsilon^2) & \frac{m_2^2(1+{\cal O}(\epsilon^2))}{2E} \\
                        \end{array}
                      \right) \left(\begin{array}{cc}  U^{\dagger}_2 & \\  & 1_2  \end{array}\right) . \label{H2+2}
\end{equation}
Here, $M_{\nu}$ is given by Eq.~(\ref{M_2+2}) while $U_2$ and $1_2$ are defined in Eq.~(\ref{U_2}). We define $\Delta m^2\equiv m_2^2-m_1^2$. Note that a diag$(U_2,1_2)$ rotation expresses the Hamiltonian in an active--sterile mass-squared basis where the top two components consist of two mass-squared states that are mostly active, while the bottom two components consist of two mass-squared states that are mostly sterile. The Hamiltonian is block diagonal in this basis in the limit $\epsilon_{1,2,3}\to 0$. In this limit, of course, we are left with the standard MSW Hamiltonian for two active neutrino flavors.

\begin{figure}
\includegraphics[width=0.9\textwidth]{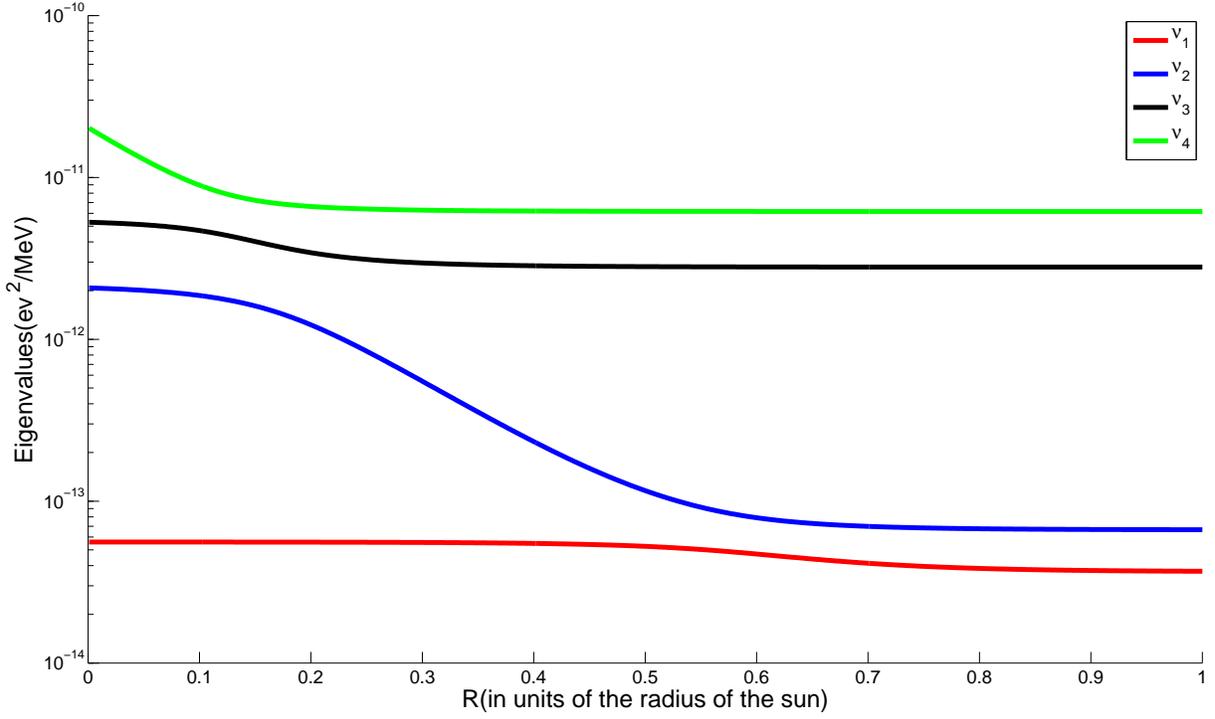}
\caption{The evolution of instantaneous eigenvalues of the Hamiltonian inside the sun  in the 2+2 scenario, Eq.~(\ref{H2+2}), for $E=10$~MeV,
$m_1^2=1\times 10^{-6}$~eV$^2$, 
$m_2^2=8.2\times 10^{-5}$~eV$^2$,
$\sin^2\theta=0.32$,
$\epsilon_1=3\times 10^{-1}$,
$\epsilon_2=4\times 10^{-1}$, and 
$\epsilon_{m3}=10^{-3}$~eV.}
\label{eigenvalues_2+2}
\end{figure}

Figure~\ref{eigenvalues_2+2} depicts the evolution of the instantaneous eigenvalues of $H$ inside the sun, as a function $R$, the distance from the sun's center, for $\epsilon_{1,2,3}\neq 0$. For illustrative purposes, we choose $E=10$~MeV,
$m_1^2=1\times 10^{-6}$~eV$^2$, 
$m_2^2=8.2\times 10^{-5}$~eV$^2$,
$\sin^2\theta=0.32$,
$\epsilon_1=3\times 10^{-1}$,
$\epsilon_2=4\times 10^{-1}$, and 
$\epsilon_{m3}=10^{-3}$~eV for the Hamiltonian parameters. The LMA MSW resonance (between the largest eigenvalue and the smallest one) can be readily identified (at $R\sim0.1$). One can also see the impact of $\epsilon_{1,2,3}\neq 0$: the top two and the bottom two eigenvalues are ``split'' for all values of  $R$. Similar to the 2+1 case discussed in the previous subsection, there are two other resonances \cite{Alex} due to the small but non-zero $\epsilon_{1,2,3}$ at larger $R$ between the two largest and two smallest Hamiltonian eigenstates. From the Hamiltonian Eq.~(\ref{H2+2}), it is easy to see that these are governed by the small parameters $\epsilon_1$ (resonance between the two lightest states) and $\epsilon_2$ (resonance between the two heaviest states). $\epsilon_3$ connects the {\sl heaviest} ``active'' state to the {\sl lightest} ``sterile'' state (and vice-versa) and, for physically interesting values of the neutrino oscillation parameters, its effects are not visible. 

As in the 2+1 case, $\Delta m^2=(7.59\pm 0.21)\times 10^{-5}$~eV$^2$ is fixed by the results of the KamLAND experiment, and $\sin^2\theta$ is constrained to be large. In this case, electron neutrino production in the sun's core and subsequent propagation can be described as follows. The electron is produced as an incoherent superposition of the two mostly active states (this is true as long as $A\gg \epsilon_{1,3}m_1^2/(2E)$ for all values of $E$ of interest). The two components then evolve adiabatically until they hit the resonances governed by $\epsilon_1$ and $\epsilon_2$. Hence the electron neutrino exits the sun as an incoherent mixture of $\sqrt{1-P_{c1}}\nu_1-\sqrt{P_{c1}}\nu_{1'}$ (with probability $\cos^2\theta_M$) and $\sqrt{1-P_{c2}}\nu_2-\sqrt{P_{c2}}\nu_{2'}$ (with probability $\sin^2\theta_M$).\footnote{Relative phases between $\nu_i$ and $\nu_{i'}$, $i=1,2$ have been ignored as neither of them is observable in practice.} Here, $\theta_M$ is defined by Eq.~(\ref{theta_M}). The crossing probabilities are given by
\begin{equation}
    P_{c,i}=\frac{e^{-\gamma_i|V_{s_ii}|^2}-e^{-\gamma_i}}{1-e^{-\gamma}},~~~~
    \gamma_i\simeq9.8\left(\frac{\epsilon_i}{10^{-4}}\right)\left(\frac{m_i^2}{8\times10^{-5}~\rm eV^2}\right)
               \left(\frac{\rm 0.862~MeV}{E}\right),~~i=1,2, \label{Pci}
\end{equation}
where $V_{s_ii}$ are given in Eq.~(\ref{V4x4}).
On their way to the surface of the earth, both of these states can undergo vacuum oscillations driven by the mass-squared differences $2m_1^2\epsilon_1$ and $2m_2^2\epsilon_2$ [see Eq.~(\ref{long_2x2})]. One can write
\begin{equation}
P_{ee}=\cos^2\theta_M\times P(\sqrt{1-P_{c1}}\nu_1-\sqrt{P_{c1}}\nu_{1'}\to \nu_e) + \sin^2\theta_M \times P(\sqrt{1-P_{c2}}\nu_2-\sqrt{P_{c2}}\nu_{2'}\to\nu_e), 
\end{equation}
where
$P(\sqrt{1-P_{c1}}\nu_1-\sqrt{P_{c1}}\nu_{1'}\to \nu_e)$ is given by Eq.~(\ref{Pmix_e}) with $P_c,m,\epsilon$ replaced by $P_{c1},m_1,\epsilon_1$, respectively, and $U_{e2,e2'}$ replaced by  $U_{e1,e1'}$, respectively. Similarly, $P(\sqrt{1-P_{c2}}\nu_2-\sqrt{P_{c2}}\nu_{2'}\to\nu_e)$ is given by Eq.~(\ref{Pmix_e}) with $P_c,m,\epsilon$ replaced by $P_{c2},m_2,\epsilon_2$, respectively. After the dust settles, 
\begin{eqnarray}
P_{ee}&=&\cos^2\theta_M\frac{\cos^2\theta}{2}\left[1+2\sqrt{P_{c1}(1-P_{c1})}\cos\left(\frac{m_1^2\epsilon_1 L}{E}\right)\right]+\sin^2\theta_M\frac{\sin^2\theta}{2}\left[1+2\sqrt{P_{c2}(1-P_{c2})}\cos\left(\frac{m_2^2\epsilon_2 L}{E}\right)\right], \nonumber \\
P_{ea}&=&\cos^2\theta_M\frac{\sin^2\theta}{2}\left[1+2\sqrt{P_{c1}(1-P_{c1})}\cos\left(\frac{m_1^2\epsilon_1 L}{E}\right)\right]+\sin^2\theta_M\frac{\cos^2\theta}{2}\left[1+2\sqrt{P_{c2}(1-P_{c2})}\cos\left(\frac{m_2^2\epsilon_2 L}{E}\right)\right], \nonumber \\
P_{es}&=&\cos^2\theta_M\frac{1}{2}\left[1-2\sqrt{P_{c1}(1-P_{c1})}\cos\left(\frac{m_1^2\epsilon_1 L}{E}\right)\right]+\sin^2\theta_M\frac{1}{2}\left[1-2\sqrt{P_{c2}(1-P_{c2})}\cos\left(\frac{m_2^2\epsilon_2 L}{E}\right)\right], \label{P_sun_2+2}
\end{eqnarray}
 where $P_{es}\equiv P_{es_1}+P_{es_2}$ and $\theta_M$ is given by Eq.~(\ref{theta_M}). In all expressions above, we have assumed all $\epsilon_{1,2,3}$ much smaller than 1. This will be justified {\it a posteriori} (see Sec.~\ref{sub:exp}) for $\epsilon_1$ and $\epsilon_2$ and we postulate that it is valid for $\epsilon_3$. The impact of a ``large'' $\epsilon_3$ is mostly felt in the diagonalization of $H$ [Eq.~(\ref{H2+2})] in the production region, which we assumed is properly described only by the matter mixing angle $\theta_M$ [see Eq.~(\ref{V_matter})]. The bound one would obtain on $\epsilon_{m3}$ via these effects is much worse (by many orders of magnitude) than the bounds on $\epsilon_{m1}$ and $\epsilon_{m2}$ discussed in the next subsection. As explained earlier, we are interested in the most stringent bound on any element of $\epsilon_{m}$ and will not consider these $\epsilon_3$ effects henceforth. On a related note, Eqs.~(\ref{P_sun_2+2}) do not depend on $\epsilon_3$, similar to Eq.~(\ref{long_2x2}). For this reason we will not discuss $\epsilon_3$ effects or bounds.

\begin{figure}
\includegraphics[width=0.6\textwidth]{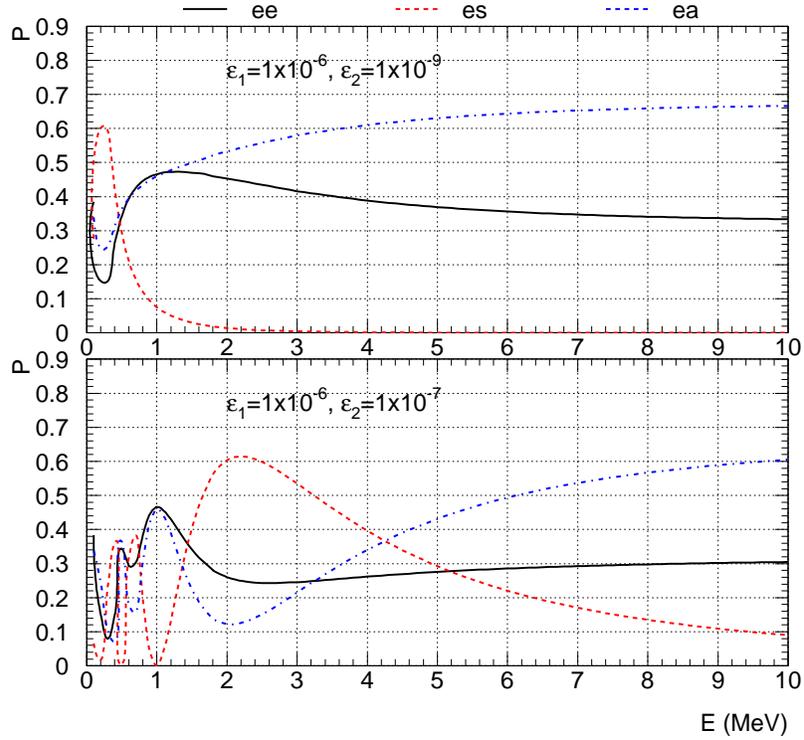}
\caption{$P_{e\zeta}$ for solar neutrinos at the surface of the earth, $\zeta=e,a,s$, as a function of neutrino energy for $m_1^2=1.0\times10^{-6}$~eV$^2$, $m_2^2=8.2\times10^{-5}$~eV$^2$, $\sin^2\theta=0.31$. TOP:  $\epsilon_1=1\times 10^{-6}$, $\epsilon_2=1\times 10^{-9}$, BOTTOM:  $\epsilon_1=1\times 10^{-6}$, $\epsilon_2=1\times 10^{-7}$.}
\label{fig:P_2+2}
\end{figure}

Figure~\ref{fig:P_2+2} depicts $P_{e\zeta}$ for solar neutrinos at the surface of the earth, $\zeta=e,a,s$, as a function of neutrino energy for $m_1^2=1.0\times10^{-6}$~eV$^2$, $m_2^2=8.2\times10^{-5}$~eV$^2$, $\sin^2\theta=0.31$ and $\epsilon_1=1\times 10^{-6}$, plus two different values of $\epsilon_2=1\times 10^{-9}$ and $\epsilon_2=1\times 10^{-7}$.  It is easy to see that dominant $\epsilon_1$ effects are most visible for low energy solar neutrinos, when $\cos^2\theta_M$ is significantly different from zero [Fig.~\ref{fig:P_2+2}(TOP)]. On the other hand, dominant $\epsilon_2$ effects are identical to the ones observed in the 2+1 case, where $\epsilon_2$ plays the role of $\epsilon$. The main distinction is the fact that in the 2+1 case $m^2$ was constrained by KamLAND, while in the 2+2 case neither $m_1^2$ nor $m_2^2$ are strongly constrained by data, except through their difference: $m_2^2-m_1^2=7.6\times 10^{-5}$~eV$^2$. In the case when both $\epsilon_1$ and $\epsilon_2$ are non-negligible [Fig.~\ref{fig:P_2+2}(BOTTOM)], one can see that $\epsilon_2$ effects are dominant for high energy solar neutrinos ({\it cf.}  Fig.~\ref{fig:P_2+1}), while $\epsilon_1$ and $\epsilon_2$
 effects ``add up'' for the low energy solar neutrinos. 

\subsection{ Constraints from Experimental Data}
\label{sub:exp}

The flux of electron neutrinos from the sun has been measured, sometimes as a function of energy, by a variety of experiments and in a variety of ways. The flux of active neutrinos from the sun has also been unambiguously measured for $^8$B neutrinos. Here, we use  recent data from \cite{SNO,SNO1,Aharmim:2008kc,SuperK,GNO,Chlorine,Borexino} in order to constrain $\epsilon$ (in the 2+1 case) and $\epsilon_{1,2}$ (in the 2+2 case). In order to compute the expected event rates at the different experiments we use parameters from the BS05(OP)  version of the Standard Solar Model (SSM) \cite{SSM05}. 

We do not attempt an exhaustive analysis of all neutrino data but instead concentrate on the features that most impact our results. We make use of the following observables. More details are provided in the Appendix. 
\begin{itemize}
\item $\phi^{CC}$(SNO), the integrate $^8$B neutrino flux measured via $\nu_e+^2$H~$\to e^-+p+p$ at SNO \cite{SNO,SNO1,Aharmim:2008kc}.  This observable is sensitive only to the electron neutrino component of the solar neutrino flux.
\item $\phi^{NC}$(SNO), the integrated $^8$B neutrino flux measured via $\nu_{e,a}+^2$H~$\to \nu_{e,a}+n+p$ at SNO \cite{SNO,SNO1,Aharmim:2008kc}. This observable is sensitive to the active neutrino component of the solar neutrino flux. 
\item $\phi^{ES}$(SuperK), the $^8$B neutrino flux measured via $\nu_{e,a}+e^-\to\nu_{e,a}+e^-$ at SuperKamiokande \cite{SuperK} integrated over a handful of recoil electron energy bins. This observable is sensitive to the active neutrino component of the solar neutrino flux. The cross section for muon or tau-type neutrino--electron scattering is about 0.16 times that of electron-type neutrino--electron scattering when integrated over the $^8$B neutrino energy spectrum. SNO also has data on neutrino--electron elastic scattering (also included), but with significantly less statistics than SuperKamiokande.
\item $\phi^{ES}$(Borexino), the $^7$Be neutrino flux\footnote{Borexino is also sensitive to neutrinos produced in the CNO cycle. These account for around 10\% of the neutrinos with energies around those of the  $^7$Be neutrinos and have been neglected. The current Borexino data is consistent with a vanishing CNO neutrino flux \cite{Borexino}.} measured via $\nu_{e,a}+e^-\to\nu_{e,a}+e^-$ at Borexino \cite{Borexino}. This observable is sensitive to the active neutrino component of the solar neutrino flux. The cross section for muon or tau-type neutrino--electron scattering is about 0.21 times that of electron-type neutrino--electron scattering for $^7$Be neutrinos.
\item $\phi^{Ga}$, the measurement of the solar neutrino flux using inverse $\beta$-decay in Gallium \cite{GNO}. This observable is sensitive to neutrinos produced in all distinct neutrino-producing fusion reactions, but only to the electron neutrino component of the solar neutrino flux.
\item $\phi^{Cl}$, the measurement of the solar neutrino flux using inverse $\beta$-decay in Chlorine \cite{Chlorine}. This observable is sensitive to neutrinos produced in all distinct neutrino-producing fusion reactions, except those involving the $p+p$ reaction, but only to the electron neutrino component of the solar neutrino flux.
\end{itemize} 
All flux measurements are defined assuming that the incoming neutrinos are all electron neutrinos.  For example, 
\begin{equation}
\phi^{ES}({\rm Borexino})=\frac{\phi_{^7\rm Be}\left(\sigma_{\nu_ee}P_{ee}+\sigma_{\nu_ae}P_{ea}\right)}{\sigma_{\nu_ee}}=\phi_{^7\rm Be}\left(P_{ee}+\frac{\sigma_{\nu_ae}}{\sigma_{\nu_ee}}P_{ea}\right),
\end{equation}
where $\phi_{^7\rm Be}$ is the SSM expectation for the $^7$Be solar neutrino flux and $\sigma_{\nu_{\zeta}e}$ is the cross section for elastic $\zeta$-type neutrino--electron scattering, $\zeta=e,a$.

Before presenting the results of a $\chi^2$ fit to the data spelled out above, it is illustrative to describe the dominant aspects of the data that constrain the presence of the extra mass-squared differences. For $^8$B neutrinos, the combined SNO and Super-Kamiokande data not only reveal that $P_{ee}\sim 0.3$ for neutrino energies above a few MeV, but also reveal that $P_{ea}\sim 0.7$. A more detailed analysis, presented in the Appendix, reveals that for $^8$B neutrinos, $P_{es}<0.37$ at the 3$\sigma$ confidence level. As advertised, this rules out at least $\epsilon\gtrsim 10^{-3}$ in the 2+1 case [see Eq.~(\ref{Pes_large_e})]. Upon further scrutiny, it is easy to see that, for $\epsilon\gtrsim 10^{-6}$, $P_{es}\sim 0.5$ in the 2+1 case, which is ruled out by the current data.

For solar neutrino energies below 1~MeV or so, information regarding neutrino oscillations is dominated by Borexino, the Homestake experiment, and the Gallium experiments. The Borexino experiment measures, with good precision, mostly for $^7$Be solar neutrinos ($E_{\nu}=0.862$~MeV), $P_{ee}=0.56\pm 0.10$ in the limit $P_{es}\to0$ \cite{Borexino}. For larger values of $P_{es}$, $P_{ee}$ is driven towards higher values in order to compensate for the depletion of the $\nu_a$ component of the $^7$Be solar neutrino flux at the earth.  This effect is depicted in more detail in the Appendix. For lower energy neutrinos, the Gallium experiments, taking into account the data, mostly, from Borexino, SNO, and SuperKamiokande, constrain $P_{ee}\sim 0.55$ for $pp$ neutrinos, and are insensitive to whether those convert into active or sterile neutrinos.\footnote{The bound on $P_{ee}$ depends indirectly on $P_{es}$ for higher energy neutrinos as these modify the measurement of $P_{ee}$ for the high energy neutrinos and hence the extraction of $P_{ee}$ for $pp$ neutrinos from the Gallium data. Details are provided in the Appendix.} 

Figure \ref{fig:chisq_2+1} depicts $\Delta\chi^2=\chi^2-\chi^2(\epsilon=0)$ as a function of $\epsilon$, in the case of two active and one sterile neutrino, for $m^2=7.6\times10^{-5}$~eV$^2$ and $\sin^2\theta=0.31$. $\chi^2$ is the result of a $\chi^2$ fit to the data described above, and we will use it in order to establish an upper bound on $\epsilon$. More specifically, we will state that values of $\epsilon$ associated to $\Delta\chi^2>4,9$ are ruled out at the two, three sigma level. While we quote a bound for a fixed value of $m^2,\sin^2\theta$, we have verified that a very similar bound is obtained for different values of these oscillation parameters (in which case we also include the measurement of $m^2$ from the KamLAND experiment in the $\chi^2$ function). We note, for example, that there is no good fit for values of $\theta$ outside the currently best fit region for $\sin^2\theta_{12}$ \cite{:2008ee}, even if one considers significant values of $\epsilon$. From the figure, we estimate that the largest value of $\epsilon$ allowed at the three sigma level is $\epsilon<2.0\times 10^{-7}$, while at the two sigma level $\epsilon<1.2\times 10^{-7}$. Furthermore, the region $2.4\times 10^{-8}<\epsilon<7.1\times 10^{-8}$ is ruled out at the two sigma level. In this region $P_{es}$ is large for solar neutrino energies around 1~MeV, while $P_{ee}$ is less than one half, in contradiction with the Borexino data. For slightly larger values of $\epsilon$, the peak in $P_{es}$ moves to solar neutrino energies larger than  1~MeV but smaller than the threshold for the SNO and SuperKamiokande experiments, where there is virtually no experimental information. In this case, the fit to the data is just as good as the one obtained for $\epsilon=0$. 
\begin{figure}
\includegraphics[width=0.6\textwidth]{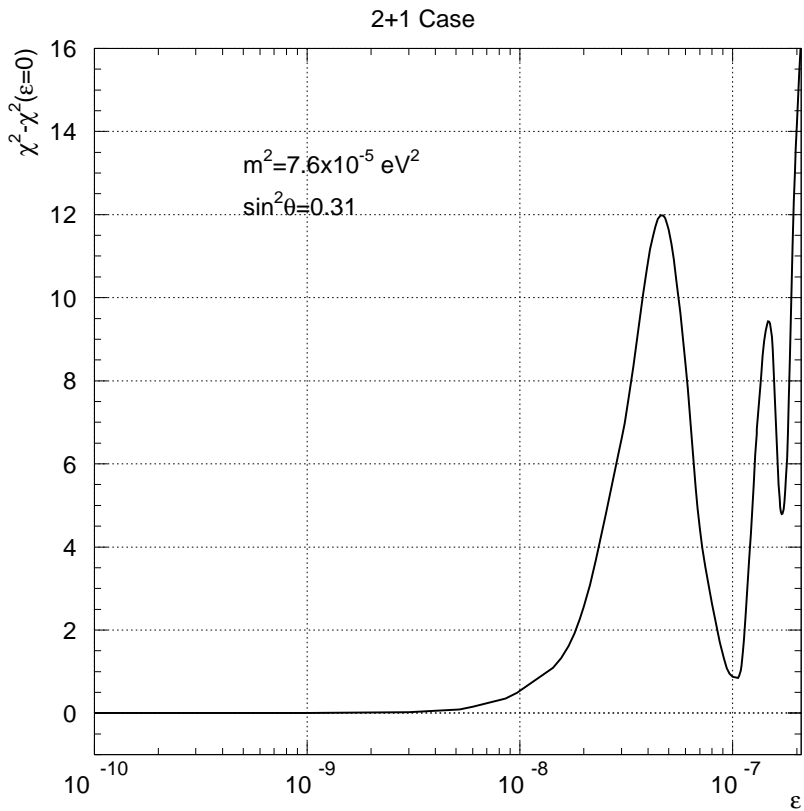}
\caption{$\chi^2-\chi^2(\epsilon=0)$ as a function of $\epsilon$ for $m^2=7.6\times 10^{-5}$~eV$^2$, $\sin^2\theta=0.31$, in the 2+1 case.}
\label{fig:chisq_2+1}
\end{figure}

It is also worthwhile pointing out that, for $\epsilon\sim 1\times 10^{-7}$, $P_{ee}$ for $^8$B neutrinos is expected to be mostly constant as a function of energy, contrary to the standard LMA case, where a slight decrease of $P_{ee}$ as a function of energy is expected. This behavior is observed for the appropriate parameter choices in other ``new physics'' scenarios including non-standard neutrino interactions \cite{NSNI}, mass-varying neutrinos \cite{Mavans}, and the introduction of light sterile neutrinos that mix very weakly with active ones \cite{deHolanda:2003tx}. Such a behavior was advertised as solution to the fact that the $^7$Be electron neutrino flux, extracted using the Chlorine data, was lower than the one predicted by the canonical LMA solution to the solar neutrino puzzle. This tension in the data has been significantly relaxed with the introduction of those from Borexino. 

 The bounds above translate into $2m^2\epsilon<(1.8,3.0)\times 10^{-11}$~eV$^2$ as the two, three sigma upper bound on the induced small mass-squared difference and $\epsilon_m<(1.0,1.7)\times 10^{-9}$~eV at the two, three sigma level.  Note that our bounds agree qualitatively with those obtained, under slightly different circumstances and after some re-interpretation of the relevant observables, in \cite{Cirelli:2004cz}. The analysis in \cite{Cirelli:2004cz} did not include results from Borexino, not available before 2008. 

In the 2+2 case, we wish to place bounds on both $\epsilon_1$ and $\epsilon_2$, and discuss which between $\epsilon_{m1}=m_1\epsilon_1$ and $\epsilon_{m2}=m_2\epsilon_2$ is more severely constrained. Unlike the 2+1 case, the data (mostly from KamLAND) do not specify $m_1^2$ or $m_2^2$, but constrain $m^2_2-m_1^2=7.6\times 10^{-5}$~eV$^2$. As discussed in the previous subsection, the solar data constrain $m_1^2\epsilon_1$ and $m_2^2\epsilon_2$. Keeping all this in mind, we will first discuss the bounds of $\epsilon_1$ and $\epsilon_2$ for fixed $m_1^2=1.0\times 10^{-6}$~eV$^2$ and $m_2^2=7.7\times 10^{-5}$~eV$^2$ and then discuss the bounds on $\epsilon_m$ for various values of $m_1^2,m_2^2$ by varying these two masses while maintaining $m_2^2-m_1^2$ constant and in agreement with the KamLAND data. As in the 2+1 case, we will discuss bounds for fixed $m_2^2-m_1^2=7.6\times 10^{-5}$~eV$^2$ and $\sin^2\theta=0.31$. We have checked that a very similar bound is obtained when one  allows these two parameters to vary in the fit. 

In more detail, we discuss the bound on $\epsilon_1$ in the limit $\epsilon_2$ very small and vice-versa. The reason for doing this is that $\epsilon_1\times \epsilon_2$ correlated effects are not very large. In the limit where $\epsilon_1$ effects are negligible, $\epsilon_2$ bounds are similar to those on $\epsilon$ in the 2+1 case. In more detail, $2m_2^2\epsilon_2<3.0~\times 10^{-11}$~eV$^2$ at the three sigma level. In the limit where $\epsilon_2$ effects are negligible, we obtain, following a procedure identical to the one discussed in the 2+1 case above, $\epsilon_1<5.3\times 10^{-7}$ at the two sigma level and $\epsilon_1<8.8\times 10^{-7}$ at the three sigma level for $m_1^2=1.0\times 10^{-6}$~eV$^2$. This translates into a new mass-squared difference $2m_1^2\epsilon_1<(1.1,1.8)\times 10^{-12}$~eV$^2$ at the two, three sigma level.

We are interested in upper bounds for the elements of $\epsilon_m$. From the bounds on the mass-squared differences above, we obtain, at the three sigma level,
\begin{eqnarray}
&\epsilon_{m1}<8.8\times 10^{-10}~{\rm eV}\left(\frac{10^{-3}~\rm eV}{m_1}\right), \label{em1_bound}\\
&\epsilon_{m2}<1.7\times 10^{-9}~{\rm eV}\left(\frac{8.8\times 10^{-3}~\rm eV}{m_2}\right). \label{em2_bound}
\end{eqnarray}
These bounds are obtained for $\sin^2\theta=0.31$ and $m_2^2-m_1^2=7.6\times 10^{-5}$~eV$^2$ and are hence correlated. In the case of an inverted mass hierarchy, for example, if $m_1=50.000\times 10^{-3}$~eV while  $m_2=50.076\times 10^{-3}$~eV, the three sigma bounds above translate into $\epsilon_{m1}<1.8\times 10^{-11}$~eV and $\epsilon_{m2}<3.0\times 10^{-10}$~eV. On the flip side, while $m_2^2>7.6\times 10^{-5}$~eV$^2$ is bounded from below, $m_1$ can be arbitrarily small, which indicates that the upper bound on $\epsilon_{m1}$, while naively stronger than that on $\epsilon_{m2}$, could be significantly worse.\footnote{Strictly speaking, some of the approximations made in Sec.~\ref{sec:model} fail when $m_1$ is of order $\epsilon_{m1}$ and hence the bound Eq.~(\ref{em1_bound}) is not applicable when $m_1\lesssim 10^{-6}$~eV. In this case however, the generic upper bound on $\epsilon_m$, as discussed in the introduction to Sec.~\ref{sec:solar} is dominated by Eq.~(\ref{em2_bound}). For even smaller values of $m_1\lesssim 10^{-9}$~eV we anticipate that there are no relevant effects due to the fact that our approximations fail. On the contrary, in this case we expect oscillations due to sterile effects to be entirely dominated by the 2+1 case discussed here.}

Before moving on to current non-solar and future constraints on $\epsilon_m$, we would like to emphasize a few relevant points. We have not considered earth matter effects while computing our oscillation probabilities, neither have we worried about large seasonal variations of the solar neutrino flux induced by the very large oscillation length induced by the new, very small mass-squared differences. The fact that we can ignore day--night effects is justified {\it a posteriori}. Earth matter effects are most significant when the matter potential of the earth is of order $\Delta m^2/2E$. For $^7$Be neutrinos ($E=0.862$~MeV) ``large'' (more than one percent) day--night effects are expected for $\Delta m^2\gtrsim 10^{-8}$~eV$^2$ (see, for example, \cite{deGouvea:1999xe}) so we estimate that, for solar neutrino energies above 100~keV (below the threshold of the Gallium experiments) large day--night effects are expected for $\Delta m^2\gtrsim 10^{-9}$~eV$^2$. This is two orders of magnitude larger than the worse upper obtained for $|m^2_2-m_{2'}^2|$. These estimates are in agreement with more detailed computations performed in \cite{Cirelli:2004cz}. Seasonal variations, the other hand, are expected for $E\lesssim 1$~MeV if the bounds on $\epsilon_m$ are saturated. These could be observed in the at, for example, the Borexino experiment. For an estimate of how large an effect one may hope to observe, we point readers to \cite{Cirelli:2004cz,deGouvea:1999wg}.

\setcounter{equation}{0} 
\setcounter{footnote}{0} 
\section{Other Consequences and Future Sensitivity}
\label{sec:future}

Whether the neutrinos are Dirac or pseudo-Dirac neutrinos is a notoriously difficult issue to resolve, especially in the limit when the right-handed neutrino mass matrix $\epsilon_m$ is tiny. Other than observing the long-wavelength oscillation driven by the new mass-squared differences between the quasi-degenerate mass-squared states, it seems virtually impossible to construct observables that will reveal the pseudo-Dirac nature of the neutrinos.

It is well-known that for  pseudo-Dirac neutrinos which arise from a seesaw Lagrangian [Eq.~(\ref{l_new})] the neutrino exchange contribution to neutrinoless double-beta decay vanishes almost perfectly \cite{0nubb_pseudodirac}. One way of understanding this is to note that the contributions of the different quasi-degenerate states to neutrinoless double-beta decay are equal and opposite, and cancel pairwise. Another way of appreciating this fact is to see that, when all neutrinos are lighter than tens of MeV, the rate for neutrinoless double-beta decay is proportional to the $ee$-element of the full neutrino mass matrix, which, as one can quickly read off Eq.~(\ref{eq:Mnu}), is exactly zero. This ``property'' is present even when the neutrinos are not pseudo-Dirac states, and depends only on all right-handed neutrino masses being small enough \cite{deGouvea:2005er,deGouvea:2006gz}. For the same reason, ``all'' (for a long list of observables see, for example, \cite{de Gouvea:2007xp}) potentially observable lepton number violating phenomena are also guaranteed to vanish almost perfectly if the only sources of lepton number violation are the Majorana neutrino masses. 

As far as kinematical probes of the neutrino masses are concerned -- the most stringent laboratory bounds come from tritium beta-decay experiments \cite{Amsler:2008zzb} -- the neutrinos behave as if they were Dirac fermions, their ``pseudo'' nature completely obscured by the very stringent constraints from oscillation experiments. The same is true for cosmological bounds on the number of neutrino species. For small enough right-handed neutrino masses, the sterile neutrinos are too weakly coupled and their states are not significantly populated in the early universe \cite{Enqvist:1990ek,Cline:1991zb,Cirelli:2004cz}. 

In the near future, current solar neutrino experiments are expected to improve our understanding of very small right-handed neutrino masses. Borexino will accumulate more data (and their systematic uncertainties are expected to go down) and improve on their results published in \cite{Borexino}. As mentioned above, Borexino is expected to measure the solar neutrino flux as a function of time, and may be sensitive to anomalous seasonal variations. The absence of such effects should help erase the local minimum around $\epsilon=1\times 10^{-7}$ in the 2+1 $\Delta \chi^2$ function, depicted in Fig.~\ref{fig:chisq_2+1}, and improve on the bound on $\epsilon$ $(\epsilon_2)$ in the 2+1 (2+2) case. It is also expected that SNO and SuperKamiokande will be able to access the low energy end ($E<5$~MeV) of the $^8$B solar neutrino spectrum and shed more light on the expected (from the canonical LMA solution) rise in $P_{ee}$ as the neutrino energy decreases.  

Future neutrino experiments are aimed at measuring with more precision (and in real time) the $pp$ solar neutrino flux and that of the $pep$ and CNO neutrinos \cite{Klein:2008zzd}. These would not only provide a more precise measurement of $P_{ee}$ for very low-energy solar neutrinos (the $pp$ neutrinos, $E\lesssim 0.5$~MeV) but would also fill the energy ``gap'' between $^7$Be neutrinos and $^8$B neutrinos (the $pep$ and CNO neutrinos, $E\sim1-2$ MeV). Indeed, Borexino is currently working on controlling the cosmogenic $^{11}$C background in order to access the $pep$ and CNO neutrino fluxes. It is expected that such results, if consistent with the canonical LMA solution to the solar neutrino puzzle, will improve, perhaps by an order of magnitude, the bounds  on $\epsilon$ in the 2+1 case or $\epsilon_{1,2}$ in the 2+2 case discussed in the previous section. 

Access to new neutrino oscillation lengths much longer than the earth--sun distance will only be provided by astrophysical neutrinos, which travel galactic and extra-galactic distances. In particular, it has been pointed out  \cite{Beacom:2003eu} that studies of the flavor composition of ultra-high energy neutrinos, to which neutrino telescopes are sensitive, are sensitive to the new mass-squared differences that characterize pseudo-Dirac neutrinos if these are as small as $10^{-16}$~eV$^2$ or perhaps $10^{-18}$~eV$^2$. Taking these estimates at face value, one would be sensitive to right-handed neutrino masses 
\begin{equation}
\epsilon_m > 1.1\times 10^{-17}\left(\frac{\delta m^2}{10^{-18}~\rm eV^2}\right)~\rm eV,
\end{equation}
where $\delta m^2$ is the mass-squared difference which can be potentially accessed by neutrino telescopes  \cite{Beacom:2003eu}. The sensitivity above is estimated in the limit where $m_3^2=2\times 10^{-3}$~eV$^2$ (in which case the neutrino mass hierarchy is normal) and is expected to be more inclusive if all neutrino masses are larger, as was observed in the 2+2 analysis in the previous section. 

\setcounter{equation}{0} 
\setcounter{footnote}{0} 
\section{Summary, Concluding Thoughts}
\label{sec:conc}

The addition of gauge singlet Weyl fermions to the Standard Model Lagrangian is a very simple and effective way of rendering the neutrinos massive, as required by experimental observations. The most general renormalizable Lagrangian consistent with this enlarged particle content, Eq.~(\ref{l_new}), while deceptively simple, provides a colorful spectrum of different phenomenological consequences, depending on the values of the new parameters in the $\nu SM$. 

For any value of the parameters $M_{ij}$ (gauge singlet Majorana mass parameters), it is possible to obtain mostly active neutrino masses in the range highlighted by experiments by properly adjusting the value of the Yukawa couplings $\lambda_{\alpha i}$. It is, hence, sufficient to discuss the different phenomenological opportunities for testing the $\nu SM$ as a function of $M_{ij}$. Further simplifying our discussion, we will consider that all $M_{ij}$ are of the same order of magnitude, and will refer to all of them as $M$. 

The largest allowed value of $M$ is of order $10^{15}$~GeV \cite{Maltoni:2000iq}. This comes from the requirement that Eq.~(\ref{l_new}) is appropriate to describe physics at energy scales of order $M$ (unitarity and perturbativity). At these high energies (and all the way down to $M$ values of order the weak scale) the only directly observable consequences of Eq.~(\ref{l_new}) are the small Majorana masses of the mostly active neutrinos. More indirectly, lepton-number violating processes involving gauge singlet fermions in the early universe may be responsible for the matter-antimatter asymmetry of the universe, through the leptogenesis mechanism \cite{Davidson:2008bu}.

\begin{figure}
\includegraphics[width=0.75\textwidth]{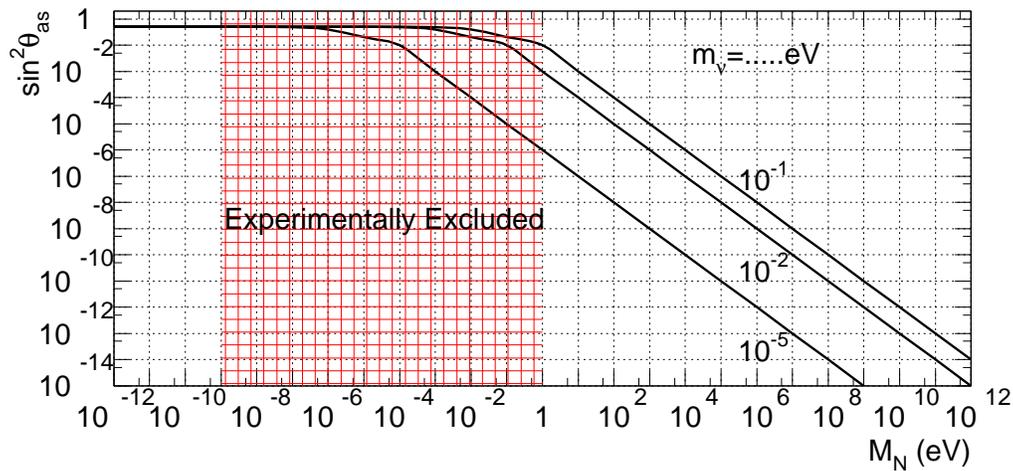}
\caption{Estimate of the magnitude of the mixing between active and sterile neutrinos $\sin^2\theta_{as}$ as a function of the right-handed neutrino mass $M_N$, for different values of the mostly active neutrino masses, $m_{\nu}=10^{-1},~10^{-2},$ and  $10^{-5}$~eV. The hatched region qualitatively indicates the values of $M_N$ that are currently excluded by the world's particle physics data.}
\label{fig:summary}
\end{figure}

For weak scale $M$ values (1~MeV$\lesssim M\lesssim 100$~GeV) the theory contains new neutral heavy leptons, which are, predominantly, gauge singlet fermions which should be produced in a variety of weak processes including charged-lepton and meson decays (see, for example, \cite{Gorbunov:2007ak,Atre:2009rg,de Gouvea:2007uz}). These neutral heavy leptons should also be produced in collider experiments and, in principle, their decays should point to lepton number violation at colliders \cite{lfv_colliders}. Indirectly, the presence of neutral heavy leptons is expected to mediate charged-lepton flavor violating phenomena and the apparent violation of universality of the weak coupling constant and of unitarity of the neutrino mixing matrix. Whether any of these phenomena is observable in practice depends on the amount of mixing between the active neutrinos $\nu_{a=e,\mu,\tau}$ and the gauge singlet fermions, $\nu_{s=s_1,s_2,\ldots}$. Fig.~\ref{fig:summary} depicts a naive estimate, as a function of $M$, of this mixing angle $\sin^2\theta_{as}$ given by Eq.~(\ref{l_new}). In the limit $M\gg m_{\nu}$,
\begin{equation}
\sin^2\theta_{as}\equiv \frac{m_{\nu}}{M},
\label{as_angle}
\end{equation}
where $m_{\nu}$ are the mostly active neutrino masses ($m_1,m_2,m_3$, constrained by oscillation experiments). The figure depicts $\sin^2\theta_{as}$ for three different values of the mostly active neutrino masses, according to  guidance provided by the current neutrino data. Experiments are sensitive to neutral heavy leptons if $\sin^2\theta_{as}\gtrsim 10^{-4}$  (best case scenario), which leads one to naively conclude that  Eq.~(\ref{l_new}), if responsible for the observed neutrino masses, cannot be tested experimentally for $M$ values in this range (1~MeV$\lesssim M\lesssim 100$~GeV). It is, however, important to emphasize that Eq.~(\ref{as_angle}) can be violated and much larger $\sin^2\theta_{as}$ values are consistent with the oscillation data \cite{de Gouvea:2007uz,Kersten:2007vk}. Whether this possibility is realized in nature can only be established experimentally. 

Sub-MeV neutral heavy leptons ((1~eV$\lesssim M\lesssim 1$~MeV) can make their presence known in astrophysics, cosmology and neutrino oscillation experiments. If $M$ values are in this range, it also turns out that, while neutrinos are Majorana fermions, the expected rates for all potentially observable lepton-number violating phenomena is strongly suppressed, as the contributions from the mostly active states cancel those from the mostly sterile ones \cite{deGouvea:2005er,deGouvea:2006gz}. In this mass range, mostly sterile neutrinos become an interesting dark matter candidate \cite{Asaka:2005an} and, at the same time, part of this mass region is only allowed if one can bypass cosmological constraints on the number of neutrino species \cite{deGouvea:2006gz}. 

In this paper, we were interested in new light neutrinos, or small values of $M$ (zero~$\le M\lesssim 1$~eV). In this case, Eq.~(\ref{as_angle}) does not apply at all, and active--sterile neutrino mixing is naively expected to be large. $M$ values of order the ``active'' neutrino masses $m_1,m_2,m_3$ are most likely excluded by failed searches for sterile neutrinos in oscillation experiments. A detailed discussion of this parameter region is very dependent on the structure of the Yukawa matrix $\lambda$ and is not the subject of this paper. Instead, we concentrated in the limit $M\ll m_1,m_2,m_3$.

If the gauge-singlet Majorana neutrino masses are much smaller than the Dirac neutrino masses $m_{\alpha i}=\lambda_{\alpha i}v$, where $v$ is the vacuum expectation value of the neutral component of the Higgs doublet, neutrinos are classified as pseudo-Dirac neutrinos. When $M_{ij}=0$ for all $i,j$, neutrinos are Dirac fermions and all experimental data can be accommodated perfectly. For $M_{ij}\neq 0$ but small enough, this conclusion remains true. Our goal was to estimate above what small but nonzero value of $M_{ij}$ is  Eq.~(\ref{l_new}) ruled out by experimental data. We find that the most stringent constraints come from the current solar neutrino data. The most conservative upper bound on the best constrained element of the gauge singlet Majorana mass matrix, in the basis defined in Sec.~\ref{sec:model}, is $M<1.7\times 10^{-9}$~eV at the three sigma confidence level. This bound is significantly stronger if the neutrino mass hierarchy is inverted or if the lightest neutrino mass ($m_1$ in the case of a normal mass hierarchy) is larger than $10^{-3}$~eV. 

The hatched region in Fig.~\ref{fig:summary} indicates the values of $M$ for which Eq.~(\ref{l_new}) is excluded as an explanation for the non-zero neutrino masses. It is remarkable that while providing a good quantitative fit to all particle physics data, the parameters of Eq.~(\ref{l_new}) are only miserably constrained.  All available neutrino data allow one to ``measure'' $M\in[0,10^{-9}]$~eV~$\cup$~$M\in[10^{-9},10^{15}]$~GeV, while the largest $\lambda_{\alpha i}\in[2\times 10^{-13},10]$. It is clear that we have only begun to properly probe the origin of neutrino masses. 

In Sec.~\ref{sec:model} we discussed a very useful basis in which to describe pseudo-Dirac neutrinos. In the basis of choice, the relation between the mass-squared splittings of the different quasi-degenerate states is directly related to the ``diagonal'' elements of $M_{ij}$, while the ``off-diagonal'' ones are only responsible, at leading order, for affecting the neutrino mixing matrix. Bounds on these off-diagonal elements were not discussed in any detail, but we argued that these are much weaker than the bounds on the diagonal elements. It is worthwhile to investigate how such effects might be probed experimentally.

\section*{Acknowledgments}

This work was performed under the auspices of the National Nuclear Security Administration of the U.S. Department of Energy at Los Alamos National Laboratory under Contract No. DE-AC52-06NA25396. It was also sponsored in part by DOE grant \# DE-FG02-91ER40684.

\appendix*
\section{Model Independent Fit to Solar Neutrino Data, Including Sterile Neutrinos}

In this appendix, we discuss a model independent, simplified fit to the solar neutrino data, along the lines of analyses performed in the past by Barger, Marfatia and Whisnant \cite{Barger:2001pf,Barger:2005si}. In a nutshell, we ``split'' the oscillation probabilities of solar neutrinos into three disjoined regions, qualitatively constrained by different experiments: 
\begin{itemize}
\item $P^H$: this is the oscillation probability of high energy solar neutrinos ($E>4$~MeV), averaged over the entire energy region. In practice, $P^H$ is dominated by the $^8$B neutrinos, and is, by far, best constrained by the real time experiments SuperKamiokande and SNO. 
\item $P^M$: this is the oscillation probability of medium energy solar neutrinos ($0.5<E<4$~MeV). In practice, constraints on this region are dominated by the monochromatic $^7$Be neutrinos ($E=0.863$~MeV). $P^M$ is best constrained by the Borexino experiment. 
\item $P^L$: this is the oscillation probability of the low energy solar neutrinos ($E<0.5$~MeV). These consist almost exclusively of $pp$ neutrinos and are only constrained by the Gallium experiments.   
\end{itemize}
Our analysis is very similar to that of \cite{Barger:2001pf,Barger:2005si}, to which we refer for details, with two exceptions. One is that the Borexino data was not available when \cite{Barger:2005si} was published and we discuss their impact here. The other is that we wish to consider that the electron neutrinos can oscillate into sterile neutrinos, and discuss how constrained is this possibility. 

\begin{figure}
\includegraphics[width=0.6\textwidth]{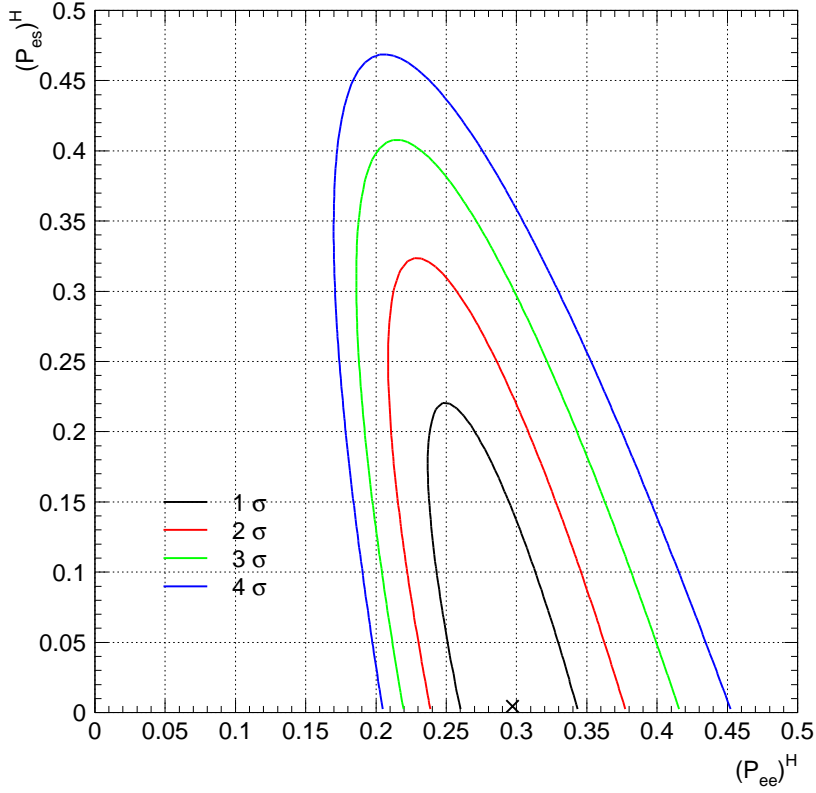}
\caption{Allowed region of the $P_{es}^H\times P_{ee}^H$ parameter space after combining data from SuperKamiokande and SNO experiments at the 1, 2, 3, and 4~$\sigma$ levels. Confidence levels are defined as regions of constant $\Delta \chi^2=2.3, 6.18, 11.83, 17.95$}
\label{P_high}
\end{figure}

Figure~\ref{P_high} depicts the allowed region of the $P_{es}^H\times P_{ee}^H$ parameter space at different confidence levels after combining data from SuperKamiokande \cite{SuperK} and SNO \cite{Aharmim:2008kc}. From the SNO experiment we learn $\phi^{CC}=(1.67\pm0.09)\times 10^{6}$~cm$^{-2}$s$^{-1}$, $\phi^{NC}=(5.54\pm0.48)\times 10^{6}$~cm$^{-2}$s$^{-1}$, while from SuperKamiokande we learn $\phi^{ES}=(2.35\pm 0.08)\times 10^{6}$~cm$^{-2}$s$^{-1}$. For the Standard Solar Model expectation we assume $\phi=(5.69\pm 0.91)\times 10^6$~cm$^{-2}$s$^{-1}$. We picked the theoretical prediction that leads to the most conservative bound on $P_{es}^H$ and assumed $P_{ee}^H+P_{es}^H+P_{ea}^H=1$. After marginalizing over $P_{ee}^H$ values, we obtain $P_{es}^H\in[0,0.37]$ at the three sigma level.

\begin{figure}
\includegraphics[width=0.6\textwidth]{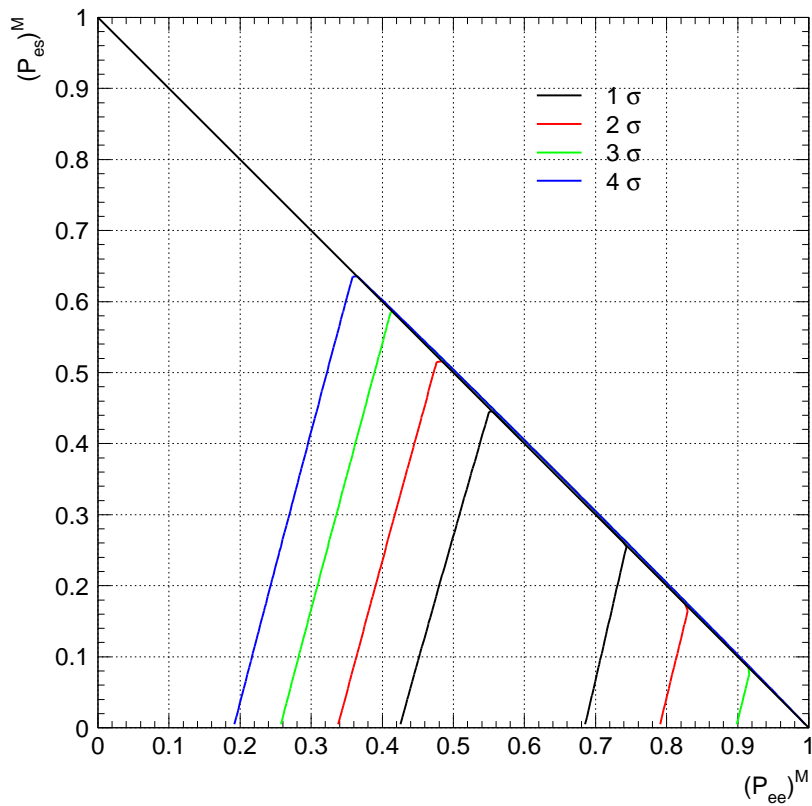}
\caption{Allowed region of the $P_{es}^M\times P_{ee}^M$ parameter space after combining data from Borexino and the Homestake experiments at the 1, 2, 3, and 4~$\sigma$ levels. Confidence levels are defined as regions of  constant $\Delta \chi^2=2.3, 6.18, 11.83, 17.95$. }
\label{P_med}
\end{figure}

Figure~\ref{P_med} depicts the allowed region of the $P_{es}^M\times P_{ee}^M$ parameter space at different confidence levels after combining data from Borexino \cite{Borexino} and the Homestake experiments \cite{Chlorine}. While computing the expected flux at Homestake, we use the value of $P_{ee}^H$ obtained in the analysis depicted in Fig.~\ref{P_high} in order to estimate the contribution of $^8$B neutrinos to the neutrino flux measured at Homestake. For more details regarding how to treat the Chlorine data, we refer readers to \cite{Barger:2001pf,Barger:2005si}. For the Borexino experiment, we learn that $49\pm 5$~counts/day/100~tons have been observed, and that $75\pm 4$~counts/day/100~tons were expected according to the Standard Solar Model. The constraint $P_{ee}^M+P_{es}^M+P_{ea}^M=1$ ``cuts off'' the part of the parameter space where $P_{ee}^M+P_{es}^M>1$. Note that our confidence level curves are defined as constant $\Delta \chi^2\equiv\chi^2(P_{ee}^M,P_{es}^M)-\chi^2_{\rm min}$ contours. One should hence be careful when translating what we refer to as an $N$ sigma ($N=1,2,3,\ldots$) bound into a probability that a certain value is allowed.  Marginalizing over $P_{ee}^M$, we get $P_{es}^M\in[0,0.57]$ at the three sigma confidence level.

It is interesting to note that, similar to the neutral current measurement at SNO and the elastic scattering result from SuperKamiokande, the Borexino result is sensitive to both electron-neutrinos and other active ($\nu_{\mu,\tau}$) neutrino species, but Borexino data alone cannot rule out $P_{ea}^M=0$. In principle, the combination of Borexino and Homestake data is sensitive to whether $P_{ea}^M\neq 0$. Alas, this is not the case in practice, as one can readily see in Fig.~\ref{P_med}.  

\begin{figure}
\includegraphics[width=0.6\textwidth]{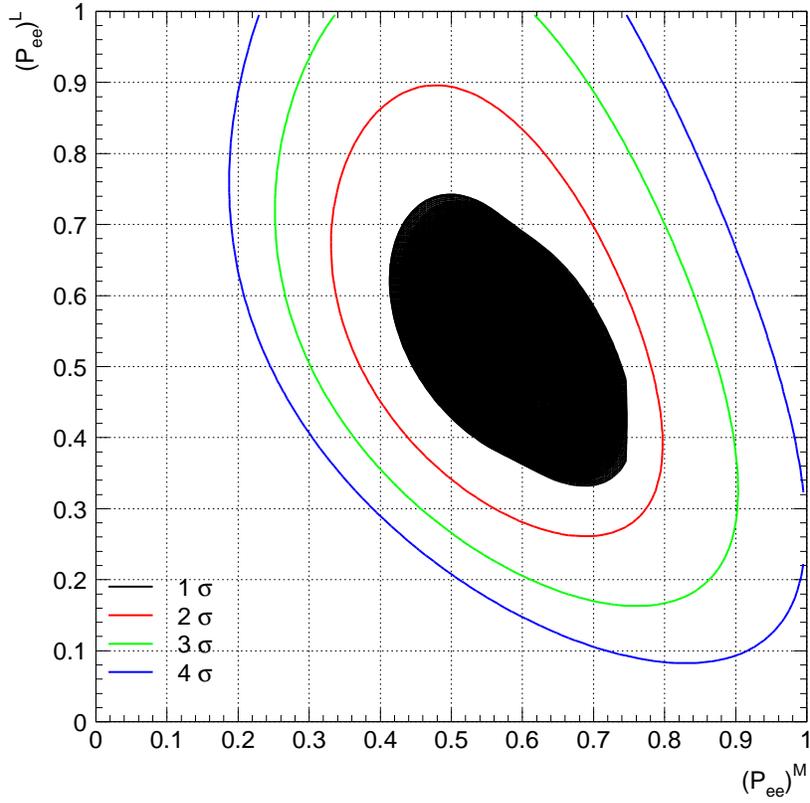}
\caption{Allowed region of the $P_{ee}^L\times P_{ee}^M$ parameter space after combining data from Borexino and the Homestake experiments at the 2, 3, and 4~$\sigma$ levels (lines), assuming $P_{es}^M=0$, and at the 1~$\sigma$ confidence level (boundary of the solid black region) marginalizing over all allowed $P_{es}^M$ values (see Fig.~\ref{P_med}). Confidence levels are defined as regions of  constant $\Delta \chi^2=2.3, 6.18, 11.83, 17.95$.
}
\label{P_low}
\end{figure}

As noted above, all information on low-energy neutrinos comes from the Gallium experiments, and these are only sensitive to $P_{ee}$ ({\it i.e.} the Gallium experiments are only sensitive to electron neutrinos from the sun).  On the other hand, since the Gallium experiments are sensitive to neutrinos of all energies above the experimental threshold, the Gallium result for $P_{ee}^L$ depends on $P_{ee}^M$ and $P_{ee}^H$. $P_{ee}^H$ is well constrained by the $^8$B experiments, so we concentrate on the interplay between $P_{ee}^L$, $P_{ee}^M$ and $P_{es}^M$ (note that there is no experimental information on $P_{es}^L$ or $P_{ea}^L$). Fig.~\ref{P_low} depicts the allowed region of the $P_{ee}^L\times P_{ee}^M$ parameter space after combining data from the Borexino and Gallium experiments (the contribution from Homestake data is also included. Its effect is very small).  In order to illustrate the effect of sterile neutrinos, the figure depicts the one sigma allowed region obtained once we marginalize over all allowed values of $P_{es}^M$ (boundary of the black region in Fig.~\ref{P_low}) and the two, three and four sigma allowed regions which are obtained if $P_{es}^M$ is set to zero in the analysis. 

Our result is consistent with that shown by the Borexino Collaboration in the Neutrino 2008 conference \cite{Borexino2008}. It also depicts the correlation between $P_{ee}^L$ and $P_{ee}^M$, courtesy of the Gallium experiments. Finally the presence of sterile neutrinos in the fit renders the allowed region in the $P_{ee}^L\times P_{ee}^M$ plane larger, but does not qualitatively invalidate the results obtained assuming that there are no sterile neutrinos.

 \end{document}